\documentclass[prd,superscriptaddress,twocolumn,nofootinbib]{revtex4}
\usepackage{amsfonts,amssymb,amsmath}
\usepackage{color}
\usepackage{graphicx}
\usepackage{dcolumn}
\usepackage{bm}
\usepackage{here}

\newcommand{\B}[1]{{\mathbb #1}}
\newcommand{\C}[1]{{\mathcal #1}}

\newcommand{\BF}[1]{{\mathbf #1}}

\newcommand{\beq}{\begin{equation}}
\newcommand{\eeq}{\end{equation}}
\newcommand{\bea}{\begin{eqnarray}}
\newcommand{\eea}{\end{eqnarray}}
\newcommand{\nn}{\nonumber}

\newcommand{\Tr}{\mathop{\rm Tr}}

\newcommand{\half}{\frac 12}
\newcommand{\third}{\frac 13}
\newcommand{\quarter}{\frac 14}
\newcommand{\eighth}{\frac 18}

\newcommand{\Slash}[1]{{\ooalign{\hfil#1\hfil\crcr\raise.167ex\hbox{/}}}}

\begin{document}

\preprint{arXiv:1011.3998 [hep-ph]}
\title{Renormalization effects on the MSSM from a calculable model of \\
a strongly coupled hidden sector}

\author{Masato Arai}
\email{Masato.Arai(AT)utef.cvut.cz}
\affiliation{Institute of Experimental and Applied Physics,
Czech Technical University in Prague, 
Horsk\' a 3a/22, 128 00 Prague 2, Czech Republic}
\author{Shinsuke Kawai}
\email{kawai(AT)skku.edu}
\affiliation{
Institute for the Early Universe (IEU),
11-1 Daehyun-dong, Seodaemun-gu, Seoul 120-750, Korea} 
\affiliation{Department of Physics, 
Sungkyunkwan University,
Suwon 440-746, Korea}
\author{Nobuchika Okada}
\email{okadan(AT)ua.edu}
\affiliation{
Department of Physics and Astronomy, 
University of Alabama, 
Tuscaloosa, AL35487, USA} 

\date{15 September 2011}

\begin{abstract}

We investigate possible renormalization effects on the low-energy mass spectrum of the minimal
supersymmetric standard model (MSSM), using a calculable model of strongly coupled
hidden sector. 
We model the hidden sector by ${\C N=2}$ supersymmetric quantum chromodynamics with 
gauge group $SU(2)\times U(1)$ and $N_f=2$ matter hypermultiplets, perturbed by a
Fayet-Iliopoulos term which breaks the supersymmetry down to  ${\C N}=0$ on a metastable
vacuum.
In the hidden sector the K\"{a}hler potential is renormalized.
Upon identifying a hidden sector modulus with the renormalization scale, and extrapolating to the
strongly coupled regime using the Seiberg-Witten solution, the contribution from 
the hidden sector to the MSSM renormalization group flows is computed.
For concreteness, we consider a model in which the renormalization effects are communicated to
the MSSM sector via gauge mediation.
In contrast to the perturbative toy examples of hidden sector renormalization studied in the 
literature,
we find that our strongly coupled model exhibits rather intricate effects on the MSSM soft scalar 
mass spectrum, depending on how the hidden sector fields are coupled to the 
messenger fields.
This model provides a concrete example in which the low-energy spectrum of MSSM particles
that are expected to be accessible in collider experiments 
is obtained using strongly coupled hidden sector dynamics.
\end{abstract}

\maketitle

\section{Introduction}
\label{sec:intro}


In the near future high-energy collider experiments such as in the Large Hadron Collider are 
expected to yield a wealth of data that help us understand physics beyond the Standard Model.
If the idea of supersymmetry (SUSY) is correct, we should detect superparticles.
Major theoretical challenges would then be to understand the mediation mechanism of SUSY 
breaking and the structure of a hidden sector, from a set of information given in the form of a
low-energy spectrum of superparticles.

The low-energy mass spectrum of a particle theory model is computed by solving
renormalization group (RG) equations.
Recently it was pointed out that the hidden sector dynamics affects RG
flows in the visible sector \cite{Dine:2004dv,Cohen:2006qc}, and thus the hidden sector can be
made visible through carefully analyzing the low-energy mass spectrum of new particles.
Based on this, RG analysis was performed for the constrained minimal supersymmetric standard model (cMSSM) \cite{Campbell:2008tt} and the gauge-mediated
supersymmetry breaking \cite{Arai:2010ds} (GMSB) scenario, both considering a simple toy 
model of hidden sector described by superpotential,
\beq
W=\frac{\lambda}{3}X^3,
\label{eqn:toyW}
\eeq
where $X$ is the hidden sector field and $\lambda$ the self-coupling.
In particular, it was found that the hidden sector effects can be crucial in determination
of the next-lightest superparticle (NLSP) \cite{Arai:2010ds},
with exciting implications in future collider experiments.

The simple model of hidden sector (\ref{eqn:toyW}) employed in these analyses 
\cite{Campbell:2008tt,Arai:2010ds}
is a mere toy, in the sense that the SUSY is assumed to be broken via some unspecified
mechanism that gives rise to a nonvanishing expectation value of the hidden sector field;
in a more complete theory a hidden sector with a spontaneous SUSY breaking mechanism needs 
to be implemented. 
In fact, a natural mechanism of SUSY breaking is believed to be a dynamical one 
\cite{Witten:1981kv}, where SUSY is broken by nonperturbative effects 
(such as instanton corrections) as the nonrenormalization theorem protects SUSY perturbatively.
It is thus important to study nonperturbative effects in the hidden sector that may affect
the RG flow of the minimal supersymmetric Standard Model (MSSM). 
The present paper is intended to provide a first step in this direction.
As it is not easy to devise a full dynamical model, we shall focus on a spontaneous SUSY 
breaking model that can be incorporated in the RG analysis of the low-energy mass spectrum.
We shall use as the hidden sector a perturbed Seiberg-Witten theory where
the duality and holomorphicity allow us to relate the strongly coupled dynamics of the 
hidden sector with the RG flow in the MSSM.

In order to discuss how the hidden sector affects the RG flow of the MSSM, we make the following
assumptions. 
We naturally identify a modulus in the hidden sector with the renormalization scale. 
Following \cite{Gildener:1976ih,Sher:1988mj}, a path is chosen from the multidimensional 
moduli space so as to parametrize a local minima of the hidden sector effective potential.
The path is then identified with the RG flow in the hidden sector, and in the Seiberg-Witten 
type models it is natural to take the paths along the singular points where the 
Bogomol'nyi-Prasad-Sommerfeld (BPS) states 
condense (see \cite{Kofman:2004yc} for discussions in a more general context).
An essential fact that we rely on in this paper is that the RG flows of the soft masses are driven 
by the wave function renormalization \cite{Giudice:1997ni,ArkaniHamed:1998kj}. 
Since the factor of the wave function renormalization is exactly calculable in Seiberg-Witten 
theory, we shall use it to discuss the RG effects in the strongly coupled region.
Thus, the effects from the hidden sector can be studied once we make reasonable assumptions 
on the hidden sector renormalization scale and the hidden-messenger coupling.

As a concrete model, we shall study a hidden sector having ${\C N}=2$ SUSY and 
$SU(2)\times U(1)$ gauge group, with $N_f=2$ matter hypermultiplets perturbed by
a Fayet-Iliopoulos (FI) term.
This theory is certainly not UV complete due to the $U(1)$ part; 
we assume that the theory is embedded in some UV complete theory and our description of the
hidden sector is valid below the scale of the Landau pole.
The effect of the FI term is to break the SUSY while the duality is maintained.
For concreteness we shall focus on the GMSB scenario \cite{GMSB} that is favored due to 
the natural suppression of flavor-changing neutral currents.
In GMSB the hidden sector fields that couple to the messengers need to be MSSM gauge singlets,
and we study two possible examples of such fields.
It turns out that our model is sufficiently nontrivial and has various intriguing features.
For example, as we will see, the hidden sector contribution to the MSSM RG equations can 
be positive or negative, depending on couplings between the hidden sector and the messenger fields; this is in contrast to the toy model studied in 
\cite{Campbell:2008tt,Arai:2010ds}, where the hidden sector effects always render the
soft scalar particles lighter.

The plan of the present paper is as follows.
In the next section we start by describing the model of the hidden sector.
In Section 3 we discuss RG flows in GMSB and, in particular, the contribution
from the hidden sector.
We present our numerical results in Section 4, and conclude in Section 5 with comments.
Technical details on the hidden sector computations based on the ${\C N}=2$ dualities,
as well as on the FI terms in the ${\C N}=2$ superspace, are collected in two Appendices.

\section{Structure of the hidden sector}

We describe in this section the hidden sector which is responsible for the SUSY breaking.
The model we shall consider is the ${\C N}=2$ supersymmetric $SU(2)\times U(1)$ gauge
theory with two matter hypermultiplets $Q$ and $\tilde Q$, perturbed by a FI term.
The model exhibits a metastable vacuum in the Coulomb branch; we use this as the
SUSY breaking sector that couples to the MSSM sector via gauge mediation.
This type of SUSY breaking scenario is particularly appealing for phenomenological applications; 
the R-symmetry is spontaneously broken and the gauginos can obtain masses.
Also, such a local minimum in a landscape is expected to be quite ubiquitous in a larger theoretical set up such as string theory.
An essential feature of the model is that the strongly coupled dynamics can be calculable due to
the ${\C N}=2$ dualities. We shall heavily rely on the machinery developed by Seiberg and Witten
\cite{Seiberg:1994rs,Seiberg:1994aj}. 
This particular type of theory has been studied extensively in \cite{
AlvarezGaume:1997fg,AlvarezGaume:1997ek,Bilal:1997st,Arai:2001pi,Arai:2007md}.

\subsection{Classical theory}
\label{sec:model-cl}

We first define our classical hidden sector Lagrangian and analyze its classical vacuum. 
The classical theory of the ${\C N}=2$  $SU(2)\times U(1)$ model was originally analyzed 
in \cite{Fayet:1975yi}.
We shall use index $i=1,2$ for the $U(1)$ and the $SU(2)$ gauge groups.
The $SU(2)$ and $U(1)$ ${\C N}=2$ vector multiplets can be written using the ${\cal N}=1$ 
vector superfields $V_i$ and the adjoint chiral superfields $A_i$, as
$(V_2, A_2)$ and $(V_1, A_1)$. 
Similarly, the matter hypermultiplets are written in terms of the ${\C N}=1$ chiral superfields as
$(Q^r_I, \tilde Q_r^I)$, where $r=1, 2$ and $I=1, 2$ are the flavor and the $SU(2)$ color indices
(the latter suppressed below).
The superfield strengths are 
$W_{i \alpha}= - \frac{1}{4} \overline{D}^2 (e^{-V_i} D_{\alpha} e^{V_i})$, with $\alpha$ the spinor index. 
The classical Lagrangian of the hidden sector is given by
\begin{eqnarray}
{\cal L}_{\rm hid}&=&{\cal L}_{\rm VM}
           +{\cal L}_{\rm  HM}
           +{\cal L}_{\rm FI} \; ,  \label{eq:lag}
\end{eqnarray}
where
\bea
{\cal L}_{\rm VM}
        &=&\frac{1}{4\pi} \mbox{Im}\left[\Tr
           \left\{\tau_{22}
           \left( \int d^4\theta A_2^\dagger e^{2 V_2} A_2 e^{-2 V_2}
           \right.\right.\right.
           \nonumber \\
 && \left.\left.\left.
 +\frac{1}{2}\int d^2 \theta  W_2^2
 \right)\right\}\right] \\
 &&+ \frac{1}{8\pi}{\rm Im}\left[\tau_{11} \left(\int d^4\theta
           A_1^\dagger A_1+\frac{1}{2} \int d^2\theta
           W_1^2 \right)\right],\nn\\
{\cal L}_{\rm HM}
   &=& \int d^4\theta  \left(
       Q_r^\dagger e^{2V_2+2V_1} Q^r
       + \tilde{Q}_r e^{-2V_2-2V_1} \tilde{Q}^{\dagger r}
          \right)  \nonumber \\
   &+& \sqrt{2} \left( \int d^2 \theta
           \tilde{Q}_r \left( A_2+A_1 \right) Q^r + h.c. \right),
   \label{cl2} 
\eea
and 
           \beq
 {\cal L}_{\rm FI}
 =\lambda \int d^2\theta A_1 + h.c.     
 \label{eq:FI}
\eeq
Our convention here is 
$\Tr (T^aT^b) = \frac{1}{2}\delta^{ab}$ for the $SU(2)$ generators $T^a$.
The complex gauge couplings of the $SU(2)$ and the $U(1)$ gauge interactions are
\beq
\tau_{22}=\frac{\theta}{\pi}+\frac{8\pi i}{g^2}, 
\qquad
\tau_{11}=\frac{8\pi i}{e^2}.
\eeq
When the FI term (\ref{eq:FI}) is absent, the theory has a global symmetry
$SU(2)_{\rm left}\times SU(2)_{\rm right}\times SU(2)_R\times U(1)$. 
The $U(1)$ charges of the hypermultiplets are normalized to be unity.
Because of the $SU(2)_R$ symmetry of the ${\C N}=2$ SUSY the FI term can be taken as a 
D-term or an F-term (see Appendix B for details). 
Here we have chosen a frame in which the FI term appears only as an F-term.
The FI term coefficient $\lambda$ is a complex number with mass dimension 2, and is assumed
to be small compared to the $SU(2)$ dynamical scale $\Lambda$;
thus the exact low-energy effective action from the ${\C N}=2$ part is unaffected by the 
additional term.
The FI term breaks the $SU(2)_R$ symmetry down to an Abelian subgroup $U(1)_{R^\prime}$, and accordingly the global symmetry of the theory is broken to
$SU(2)_{\rm left}\times SU(2)_{\rm right}\times U(1)_{R^\prime}\times U(1)_R$.

It is straightforward to find the classical potential from the above Lagrangian,
\begin{eqnarray}
 V_{\rm cl} &=&\frac{1}{g^2}\Tr [A_2, A_2^\dagger]^2
       +\frac{g^2}{2}
 ( q_r^{\dagger} T^a q^r- \tilde{q}_r T^a  \tilde{q}^{\dagger r} )^2
 \nonumber \\
    &+& q_r^\dagger [A_2, A_2^\dagger] q^r
     - \tilde{q}_r [A_2,A_2^\dagger] \tilde{q}^{\dagger r}
    +2 g^2 |\tilde{q}_r T^a q^r |^2  \nonumber \\
   &+&\frac{e^2}{2} \left(q_r^\dagger q^r- \tilde{q}_r
       \tilde{q}^{\dagger r} \right)^2
     +e^2 |\sqrt 2\tilde{q}_r q^r + \lambda |^2 \nonumber \\
   &+& 2 \left( q_r^\dagger |A_2+A_1|^2 q^r
    +\tilde{q}_r |A_2+A_1|^2 \tilde{q}^{\dagger r} \right) \; ,
\end{eqnarray}
where $A_2$, $A_1$, $q^r$ and $\tilde{q}_r$ are the scalar components
of the corresponding chiral superfields.
The minima of the potential are obtained by the stationarity condition of the potential.
In the Higgs branch the $SU(2)\times U(1)$ symmetry is completely broken and the moduli
is along 
\bea
A_1=A_2=0,
 \quad
q^1=\left(\begin{array}{c}
v\\ 0\end{array}\right), 
\quad
q^2=\left(\begin{array}{c}
0\\ v\end{array}\right), \nn\\
\tilde q_1=\left(\begin{array}{cc}
\tilde v & 0\end{array}\right), 
\quad
\tilde q_2=\left(\begin{array}{cc}
0 & \tilde v\end{array}\right).
\eea
The components refer to the $SU(2)$ color, and $v, \tilde v\in {\B C}$ satisfy
$v\tilde v=-\frac{\lambda}{2\sqrt 2}$ from the stationarity condition. 
In this branch the classical scalar potential $V_{\rm cl}$ vanishes.
Hence the SUSY is preserved in the Higgs branch.

In the presence of the FI term the Coulomb branch is lifted and the vacuum is not 
supersymmetric anymore.
A vacuum in this branch is parametrized as
\bea
A_1=a_1, 
\quad
A_2=\half\left(\begin{array}{cc}a_2 & 0 \\0 & -a_2\end{array}\right),\nn\\
q^1=\tilde q_1=\left(\begin{array}{c}v \\0\end{array}\right), \quad
q^2=\tilde q_2=0,
\eea
where $a_1, a_2, v\in{\B C}$.
One possible branch is parametrized by $z\in{\B C}$, such that 
\beq
A_2 + A_1 = \left(\begin{array}{cc}
        \frac{a_2}{2} & 0 \\  0 & -\frac{a_2}{2}  \end{array}\right)
            + \left(\begin{array}{cc} a_1 & 0 \\ 0 & a_1 \end{array}\right)
            \equiv  \left(\begin{array}{cc} 0 & 0 \\
                     0 & z  \end{array}\right). \nonumber
        \label{eq:mix}
\eeq
In this case
\beq
v^2=-\frac{2\sqrt 2 e^2\lambda}{4 e^2+g^2},
\eeq
from the stationarity conditions and the potential minima become
\begin{eqnarray}
V=\frac{|\lambda|^2 e^2 g^2}{4e^2+g^2}.
\end{eqnarray}
The SUSY is seen to be broken at tree level by the nonzero FI parameter $\lambda$.
The SUSY breaking scales are found to be
\bea
F_1&=&-\frac{e^2g^2\lambda^\dag}{4e^2+g^2},\nn\\
F_2&=&\frac{4e^2g^2\lambda^\dag}{4e^2+g^2}.
\eea

\subsection{Qunatum theory}
\label{sec:model-q}

The low-energy dynamics including the quantum effects is described by the Wilsonian 
effective action ${\C L}_{\rm eff}$.
While obtaining such an effective action by direct path integral is an arduous task, 
in the ${\C N}=2$ supersymmetric theory it is possible to determine the exact form exploiting
the duality and holomorphicity \cite{Seiberg:1994aj,Seiberg:1994rs}.
Our strategy here is to make a full use of the ${\C N}=2$ technology by treating the FI term as perturbation.
Accordingly, the perturbative parameter of mass dimension 2 is assumed to be small
compared to the $SU(2)$ dynamical scale $\Lambda$, i.e.
$\lambda\ll\Lambda^2$.
Then it is possible to expand the low-energy effective action of our theory around the ${\C N}=2$
result \cite{Arai:2001pi,Arai:2007md},
\beq
{\C L}_{\rm eff}={\C L}_{\rm SUSY}+{\C L}_{\rm soft} + {\C O}(\left(\lambda/\Lambda^2\right)^2).
\label{eqn:qL}
\eeq
Ignoring the FI term the classical $SU(2)\times U(1)$ gauge symmetry is broken to $U(1)_c\times U(1)$ in the Coulomb branch except at the origin of the moduli $(a_1, a_2)=(0,0)$.
This $U(1)_c$ is the remnant of the $SU(2)$. 
The $U(1)$ part is not asymptotically free and we restrict our moduli to be below the Landau pole,
$|a_1|\leq\Lambda_L$.  
Since the $U(1)$ and $SU(2)$ interact through coupling to the hypermultiplets, we also impose
$|a_2|\leq\Lambda_L$. 
We shall consider $\Lambda_L$ to be much larger than the $SU(2)$ dynamical scale $\Lambda$,
and regard $\Lambda_L$ as a cutoff scale of our theory.

The ${\C N}=2$ part, discussed also in \cite{Arai:2001pi,Arai:2007md} in detail,
consists of the vector multiplets and the hypermultiplets, 
${\C L}_{\rm SUSY}={\C L}_{\rm VM}+{\C L}_{\rm HM}$. 
The former contains the superfield $A_1$ for the $U(1)$
and $A_2$ for the unbroken abelian subgroup $U(1)_c$ of $SU(2)$.
The Lagrangian is
\beq
{\C L}_{\rm VM}
=\frac{1}{8\pi}{\rm Im}\left[
\int d^4\theta\frac{\partial{\C F}}{\partial A_i} A_i^\dag
+\half
\int d^2\theta
\tau_{ij}W^{\alpha i}W^j_\alpha
\right],
\label{eqn:VM}
\eeq
where ${\C F}$ is the ${\C N}=2$ prepotential depending on $A_1$, $A_2$, $\Lambda$, $\Lambda_L$, 
and 
\beq
\tau_{ij}=\frac{\partial^2{\C F}}{\partial A_i\partial A_j}.
\label{eqn:tau}
\eeq
The repeated $i, j=\{1, 2\}$ are summed over.  
The effective Lagrangian for the hypermultiplets is
\bea
{\C L}_{\rm HM}
&=&\int d^4\theta\Big(M_r^\dag e^{2n_mV_{2D}+2n_eV_2+2nV_1}M^r\nn\\
&&\quad+\tilde M_re^{-2n_mV_{2D}-2n_eV_2-2nV_1}\tilde M^{r\dag}\Big)\\
&&\hspace{-10mm}+\sqrt 2\Big\{\int d^2\theta\tilde M_r(n_mA_{2D}+n_eA_2+nA_1)M^r
+h.c.\Big\},\nn
\eea
where $V_{2D}$ and $A_{2D}$ are the dual variables of $V_2$ and $A_2$, and 
$M^r$, $\tilde M_r$ are the chiral superfields representing quarks, monopoles and dyons that become light in the vicinity of the singular points.
These BPS states have quantum numbers $(n_e, n_m)_n$, with $n_e$ and $n_m$ the electric 
and magnetic $U(1)_c$ charges, and $n$ represents the $U(1)$ charge.
Denoting the dual variables of $a_1$ and $a_2$ as 
\beq
a_{1D}\equiv\frac{\partial{\C F}}{\partial a_1},
\quad
a_{2D}\equiv\frac{\partial{\C F}}{\partial a_2},
\eeq
the masses of the BPS states are 
\beq
M_{\rm BPS}=\sqrt 2\vert n_m a_{2D}+n_e a_2+na_1\vert.
\eeq
This coincides with the spectrum of the ${\C N}=2$ supersymmetric QCD with gauge group
$SU(2)$ and $N_f=2$, with the two hypermultiplet masses $m_1=m_2=\sqrt 2 a_1$. 
The equivalence is further justified and utilized in Appendix A.
The soft term takes the same form as the classical FI term,
\beq
{\C L}_{\rm soft} = \lambda\int d^2\theta A_1 + h.c.
\eeq

\subsection{The effective potential}
\label{sec:effpot}

\begin{center}
\begin{figure*}
\includegraphics[scale=1.00]{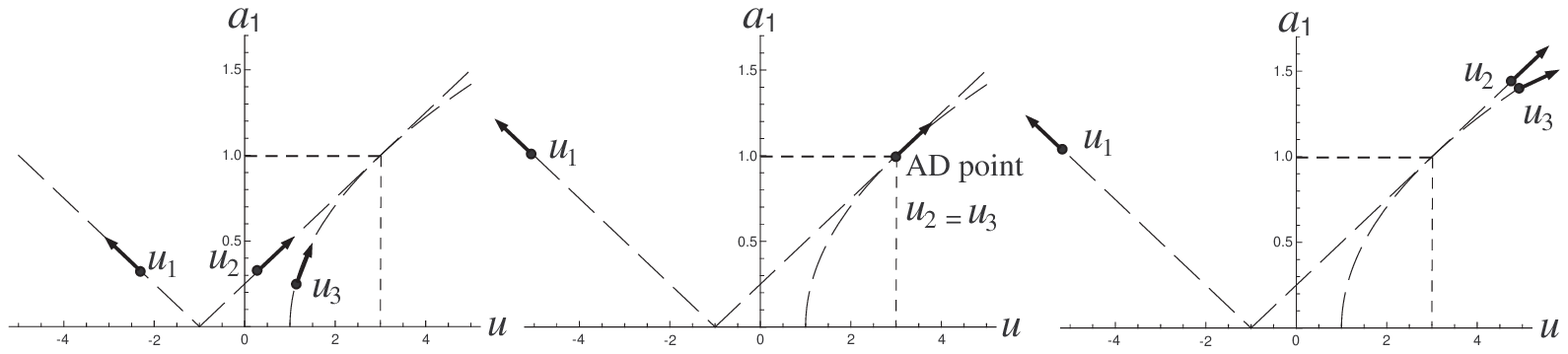}
\caption{\label{singular-t}
The position of the singular points $u_1$, $u_2$, $u_3$ in the $u$-$a_1$ plane
($u\in{\B R}$, $a_1\in{\B R}$ here and throughout the paper). 
We call $u_1$ the left dyon singular point, and $u_2$ the right dyon singular point.
$u_3$ on the left panel is a monopole singular point, while on the right panel it is a quark singular
point.
The panel in the middle shows the AD point where $u_2$ and $u_3$ coincide.
Flows are shown as $a_1$ is increased.
We have set $\Lambda=2\sqrt 2$.}
\end{figure*}
\end{center}

The effective potential obtained from the effective Lagrangian (\ref{eqn:qL}) is
\beq
V_{\rm eff}=\half b_{ij}D_iD_j+b_{ij}F_iF_j^\dag +|F_M|^2+|F_{\tilde M}|^2,
\eeq
where
\beq
b_{ij}=\frac{1}{8\pi}{\rm Im}\tau_{ij}
\label{eqn:bij}
\eeq
is the K\"{a}hler metric (computed for appropriate local variables) and
$D_i$, $F_i$, $F_M$, $F_{\tilde M}$ are the auxiliary fields
for the superfields $V_i$, $A_i$, $M$,  $\tilde M$.
The effective potential depends on $a_1$ and  $a_2$ as well as on the 
correct light degrees of freedom $M$ and $\tilde M$.
By solving the equations of motion the effective potential is written as,
\bea
V_{\rm eff}&=& S\Big\{ (M^rM_r^\dag-\tilde M_r\tilde M^{r\dag})^2+4|M^r\tilde M_r|^2\Big\}\nn\\
&&\quad+2T(M^rM_r^\dag+\tilde M_r\tilde M^{r\dag})+U\nn\\
&&\quad-\frac{\sqrt 2 b_{12}}{\det b}\Big\{\lambda M_r^\dag\tilde M^{r\dag}+ h.c.\Big\},
\eea
with
\bea
S&=& \frac{b_{11}}{2 \det b},\nn\\
T&=& |n_m a_{2D}+n_e a_2+n a_1|^2,\nn\\
U&=&\frac{b_{22}|\lambda|^2}{\det b}.
\eea
The minima of the potential are found by solving the stationarity conditions for the fields $M$ and
$\tilde M$, 
\bea
\frac{\partial V_{\rm eff}}{\partial M_r^\dag}
&=&S\left\{2(|M|^2-|\tilde M|^2)M^r+4 M^s\tilde M_s\tilde M^{r\dag}\right\}\nn\\
&&+2TM^r-\sqrt 2\lambda\frac{b_{12}}{\det b}\tilde M^{r\dag}=0,
\label{eqn:stat1}\\
\frac{\partial V_{\rm eff}}{\partial \tilde M^{r\dag}}
&=&S\left\{2(|\tilde M|^2-|M|^2)\tilde M_r+4 M^s\tilde M_sM_r^\dag\right\}\nn\\
&&+2T\tilde M_r-\sqrt 2\lambda\frac{b_{12}}{\det b}M_r^\dag=0,
\label{eqn:stat2}
\eea
where
$|M|^2=M^rM_r^\dag$ and $|\tilde M|^2=\tilde M_r\tilde M^{r\dag}$.
From (\ref{eqn:stat1}) and (\ref{eqn:stat2}), 
\beq
(|M|^2-|\tilde M|^2)\left[S(|M|^2+|\tilde M|^2)+T\right]=0,
\eeq
which implies
\beq
|M|^2=|\tilde M|^2,
\eeq 
as $T>0$ except right at the singular points, and $S>0$.
Substituting this back into (\ref{eqn:stat1}) and (\ref{eqn:stat2}), the effective potential at the
stationary points is found to be
\beq
V_{\rm min}=U-4S|M|^4.
\label{eqn:Veff}
\eeq
The second term on the right-hand side represents condensation of the light BPS degrees of freedom 
that lower the potential from $V_{\rm min}=U$ (note that $S>0$).
The expectation value of the chiral fields is also found to be
\beq
|M|^2=|\tilde M|^2=\frac{1}{2S}\left\{
\left|
\frac{\lambda b_{12}}{\sqrt 2\det b}\right|-T\right\}. \label{cond}
\eeq

\begin{figure}
\begin{center}
\includegraphics[scale=1.00]{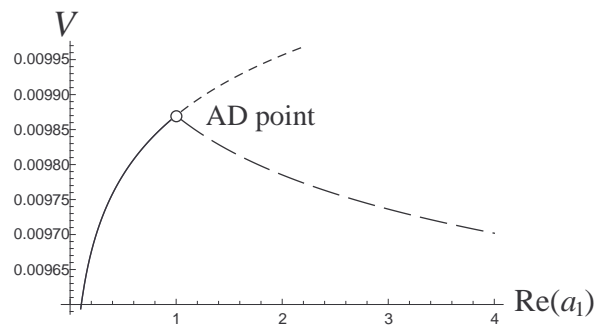}
\end{center}
\caption{\label{pott}
The local minima of the scalar potential plotted against the flows of the singular points,
for $a_1\in{\B R}$.
The solid curve shows the minima of the potential along the two dyon singular points
$u_1$ and $u_2$, which coincide for $0\leq a_1 \leq\frac{\Lambda}{2\sqrt 2}$.
$a_1=\frac{\Lambda}{2\sqrt 2}$ corresponds to the AD point.
When $a_1>\frac{\Lambda}{2\sqrt 2}$ the potential minimum for 
$u_2$ ceases to exist, 
while $u_1$ (the dotted curve) continues to be a minimum with $V_{\rm min}^1$
increasing. 
The other branch (the dashed curve) is a potential minimum around the quark singular
point $u_3$, which shows up for $a_1>\frac{\Lambda}{2\sqrt 2}$ and corresponds to a 
runaway vacuum. Here we set $\lambda=0.1$ and $\Lambda=2\sqrt 2$.
}
\end{figure}

The minima of the potential (\ref{eqn:Veff}) is a functional of the BPS fields $M, \tilde M$ as well as 
the K\"{a}hler metric (\ref{eqn:bij}), which is found by an exact method detailed in 
Appendix A.
As it turns out the theory is equivalent to ${\C N}=2$ supersymmetric QCD with gauge group 
$SU(2)$ and $N_f=2$ matter hypermultiplets having a common mass
fixed by the $U(1)$ moduli parameter as $m=\sqrt 2 a_1$.
Potential minima appear at singular points where the light BPS degrees of freedom
condense.
The location of the singular points are determined by a cubic curve \cite{Seiberg:1994aj}. 
There are three singular points apart from the one at the infinity, and they are parametrized by the
$U(1)$ and $SU(2)$ moduli parameters $a_1$ and $u=\Tr (A_2^2)$.
The three singular points $u=u_{1,2,3}$ as functions of $a_1$ and $\Lambda$ are given by
\beq
u_1=-\sqrt{2}a_1 \Lambda -\frac{\Lambda^2}{8},~~
u_2=\sqrt{2}a_1 \Lambda - \frac{\Lambda^2}{8},~~
u_3=2 a_1^2 + \frac{\Lambda^2}{8}. 
\label{eqn:s-f}
\eeq
If $u$ and $a_1$ go off the real axes, $V_{\rm eff}$ starts to increase, 
moving away from the minima. 
As we are interested in physics near minima of the potential we shall be concerned with real 
values of $u$ and $a_1$.
Numerical study shows the following structure of singularities and the corresponding vacua 
for $u, a_1\in{\B R}$
(see Figs.\ref{singular-t} and \ref{pott}; see also Fig.2 of \cite{Bilal:1997st}):

\begin{description}
  \item[{\bf (1)} $0\leq {\rm Re}(a_1)<\frac{\Lambda}{2\sqrt 2}$ ] 
  -- There are two dyon ($u_1$, $u_2$), and one monopole ($u_3$) singular points.
  The two dyon points give local potential minima ($V_{\rm min}^1$ and 
  $V_{\rm min}^2$), while the point $u_3$ does not correspond to a minimum of the potential.
  At $a_1=0$, $u_1$ (the left dyon) and $u_2$ (the right dyon) are degenerate and this type of
  vacua is analyzed in \cite{AlvarezGaume:1996zr}.
  Note that the effective theory breaks down at the $a_1=0$ point and a direct analysis is difficult.
  Nevertheless, we know that $U>0$ and it is unlikely that the instanton effects [the second term of 
  (\ref{eqn:Veff})] cancel $U$ exactly. 
  Also, SUSY is broken at the tree level and it is unlikely that it is recovered by quantum effects.
  For these reasons it is natural to suppose that the potential minimum at coincident $u_1$ 
  and $u_2$ is a SUSY breaking metastable vacuum.
  By dimensional analysis we assume $V_{\rm min}\sim\lambda^2$ at this point, which is small 
  but nonzero.
  As $a_1$ is increased, the two potential minima increase but continue to be 
  $V_{\rm min}^1=V_{\rm min}^2$ (Fig.\ref{pott}).

  \item[{\bf (2)} ${\rm Re}(a_1)=\frac{\Lambda}{2\sqrt 2}$ ]
  -- $u_2$ and $u_3$ collide at the Argyres-Douglas (AD) point. 
  The local effective Lagrangian description breaks down at this point, 
  due to overlapping of the wave functions of the BPS states.

  \item[{\bf (3)} ${\rm Re}(a_1)>\frac{\Lambda}{2\sqrt 2}$ ]
  -- There are again three singular points.
  $u_1$ and $u_2$ are dyons, while $u_3$ is now a quark singular point.
  The minimum $V_{\rm min}^1$ continues to increase, while the minimum at the quark singular
  point $V_{\rm min}^3$ starts to decrease,
  giving a runaway vacuum at $a_1$, $u\rightarrow\infty$.
  $u_2$ does not correspond to a potential minimum in this region.
\end{description}
%
%

There are two possible SUSY vacua in the model:
the first one is on the Higgs branch, and the other is the runaway vacuum on the Coulomb
branch (although it is somewhat questionable to call the latter a SUSY vacuum since the theory 
is not well defined beyond the Landau pole).
Clearly, $V_{\rm min}^3$ is almost zero in the runaway vacuum as $\Lambda_L$ is taken to be 
sufficiently large.
We approximate the runaway vacuum to be at $a_1\sim\Lambda_L$.
The decay rate of the local potential minimum at $u_1=u_2=-\Lambda^2/8$, $a_1=0$ into
these SUSY vacua is computed in \cite{Arai:2007md}.
The bounce action corresponding to the decay from the local minimum
to the runaway vacuum is computed with triangle approximation
\cite{Coleman:1977py,Callan:1977pt,Duncan:1992ai}, as
\beq
B\sim\frac{\Lambda_L^4}{\lambda^2}.
\eeq
This is extremely large as the potential barrier is very wide. 
The bounce action to the Higgs vacuum at $a_1=a_2=0$ and $q,\tilde q\sim\sqrt\lambda$ is
estimated similarly.
For  $\Lambda=2\sqrt 2$ and $\lambda=0.1$ we have $B\sim{\C O}(10^6)$.
In both decay channels the decay rate $e^{-B}$ is extremely small, and hence the 
lifetime of the false vacuum is very large. 
Thus it is regarded as a metastable vacuum.

\section{Hidden sector renormalization effects on MSSM}
\label{sec:hidden}

The spontaneous SUSY breaking in the hidden sector is communicated to the visible sector 
via messenger fields, which, in the GMSB scenario we consider, couple to the MSSM fields 
through the standard gauge and gaugino interactions. 
After integrating out the messenger fields (i.e. below the messenger scale $M_{\rm m}$), 
the interaction
between the hidden and the visible sectors is described by the Lagrangian, 
\bea
&&{\C L}_{\rm int}=k_i\int d^4\theta\frac{X^\dag X}{M_{\rm m}^2}\Phi^\dag_i\Phi_i\nn\\
&&\qquad\qquad +\left(w_a\int d^2\theta\frac{X}{M_{\rm m}}W^{a\alpha}W^a_\alpha+h.c.\right).
\label{eqn:h-v}
\eea
Here, $X$ is a MSSM gauge singlet field in the hidden sector, 
$\Phi_i$ and $W^a$ the MSSM superfields
[the indices $i=\{Q, U, D, L, E, H_u, H_d\}$ and $a=\{1,2,3\}$ denote the MSSM chiral multiplets
and the MSSM gauge groups $U(1)$, $SU(2)$, $SU(3)$, respectively].
The coupling $k_i$ is subject to wave function renormalization of the hidden sector fields,
while the nonrenormalization theorem forbids $w_a$ to be renormalized.
In our model of the hidden sector there are multiple candidates for such $X$. 
Below we shall see that different choices of $X$ lead to different types of contribution to the
MSSM RG equations.
Here and below we assume the minimal GMSB, with messenger fields belonging to 
${\BF 5}+\bar{\BF 5}$
of an $SU(5)$ with the sum of the Dynkin indices $N_5=1$.

\subsection{Gauge-mediated SUSY breaking}

Let us first describe generic features of hidden sector contributions to the MSSM
RG flow \cite{Cohen:2006qc,Campbell:2008tt}, 
focusing on the GMSB scenario \cite{Arai:2010ds}.
As SUSY is broken in the hidden sector the gauginos and scalars in the MSSM
acquire masses through 1- and 2-loop corrections, respectively
[the second and first terms in Eq. (\ref{eqn:h-v})].
The gaugino masses at the messenger scale $\mu=M_{\rm m}$ are
\beq
M_a(t=0)=\frac{\alpha_a}{4\pi}\frac{\langle F_{X}\rangle}{M_{\rm m}},
\label{eqn:GauginoMass}
\eeq
where $t=\ln ({\mu}/{M_{\rm m}})$,
$F_X$ is the F-term for the hidden sector field $X$,
$\alpha_a={g_a^2}/{4\pi}$ and $g_a$ ($a=1,2,3$) are the gauge couplings of $U(1)$, $SU(2)$, $SU(3)$.
The soft scalar masses at the messenger scale are given by
\beq
m_i^2(t=0)=2\left(\frac{\langle F_X\rangle}{M_{\rm m}}\right)
^2\sum_{a=1}^3C_2^a(R_i)\left(\frac{\alpha_a}{4\pi}\right)^2, \label{smass}
\eeq
where $C_2^a(R_i)$ are the quadratic Casimir for the matter fields in representation $R_i$
of the $a$-th MSSM gauge group.
These masses generated at the messenger scale set boundary conditions of the RG flow.
The A-terms at the messenger scale are set to be zero.

The coefficients $k_i$ of (\ref{eqn:h-v}) are renormalized by the MSSM gauge interactions and the
wave function renormalization of the hidden sector fields, $X\rightarrow Z_X^{-\half} X$.
The flow is described by the RG equations,
\beq
\frac{d}{dt}k_i(t) =\gamma(t) k_i(t)
	 -\frac{1}{16 \pi^2}\sum_{a=1}^3 8 C_2^a(R_i)g_a^6(t) G_a,
\label{eqn:RGE1}	 
\eeq
where
\beq
G_a \equiv w_aw_a^\dagger. 
\eeq
In the first term $\gamma(t)$ is the anomalous dimension at $t=\ln(\mu/M_{\rm m})$ arising 
from the hidden sector interaction (\ref{eqn:h-v}), and the second term is the leading visible 
sector contribution. 
The RG equations are solved as
\begin{eqnarray}
&& \hspace{-5mm} k_i(t)=k_i(0)\exp\left(-\int_t^0 dt^\prime \gamma(t^\prime)\right) 
\label{eqn:RGE-s1} \\
&& \hspace{-5mm} +\frac{1}{16\pi^2}\sum_{a=1}^3 8 C_2^a(R_i) \int_t^0d s g_a^6(s) 
 \exp\left(-\int_t^s dt^\prime \gamma(t^\prime) \right) G_a. \nonumber
\end{eqnarray}
The evolution of the MSSM scalar masses now includes hidden sector contributions. 
In the leading order these masses are
\begin{eqnarray}
& \displaystyle m_{i}^2(t)=k_i(t)\left(\frac{\langle F_X\rangle}{M_{\rm m}}\right)^2, 
& \label{RGE-s}
\end{eqnarray}
with 
\begin{eqnarray}
 k_i(0) = 2\sum_{a=1}^3 C_2^a(R_i)\left(\frac{\alpha_a(M_{\rm m})}{4\pi}\right)^2.
\end{eqnarray}
The gauge couplings run according to the standard 1-loop formula,
\bea
&&\hspace{-10mm}
\frac{1}{\alpha_a(\mu)}\nn\\
&&\hspace{-10mm}
=\left\{
\begin{array}{ll}
\displaystyle
\frac{1}{\alpha_a(M_Z)}-\frac{b_a}{2\pi}\ln\frac{M_S}{M_Z}
\displaystyle
-\frac{b_a^S}{2\pi} \ln\frac{\mu }{M_S}, 
&(\mu > M_S) \\
\displaystyle
\frac{1}{\alpha_a(M_Z)}-\frac{b_a}{2\pi}\ln\frac{\mu}{M_Z}, 
&(\mu \le M_S)
\end{array}
\right.
\eea
where 
$(b_1^S, b_2^S, b_3^S)=(-3, 1, 33/5)$ for the MSSM
and $(b_1, b_2, b_3)=(-7, -19/6, 41/10)$ for the Standard Model gauge couplings.
$M_Z$ and $M_{\rm S}$ are the $Z$-boson mass and a typical soft mass scale, 
respectively.

The first term in (\ref{eqn:RGE1}) implies that the hidden sector renormalization contributes to
the RG equations for the MSSM soft scalar masses as
\beq
\frac{dm_i^2}{dt}=\left. \frac{dm_i^2}{dt}\right\vert_{\rm MSSM} +\gamma(t) m_i^2.
\eeq
Explicitly, the RG equations are
\begin{widetext}
\bea
8\pi^2\frac{d m^2_Q}{dt}&=&\xi_t+\xi_b-\frac{16}{3}g_3^2 M_3^2-3g_2^2M_2^2
-\frac{1}{15}g_1^2 M_1^2
+\frac 15 g_1^2 \xi_1 + 8\pi^2 \gamma m_Q^2,
\label{eqn:RGE_Q}\\
8\pi^2\frac{d m^2_U}{dt}&=&2\xi_t-\frac{16}{3}g_3^2 M_3^2-\frac{16}{15}g_1^2 M_1^2
-\frac 45 g_1^2\xi_1+8\pi^2 \gamma m_U^2,\\
%
%
8\pi^2\frac{d m^2_D}{dt}&=&2\xi_b-\frac{16}{3}g_3^2 M_3^2-\frac{4}{15}g_1^2 M_1^2
+\frac 25 g_1^2\xi_1+8\pi^2 \gamma m_D^2,\\
8\pi^2\frac{d m^2_L}{dt}&=&\xi_\tau-3 g_2^2 M_2^2-\frac{3}{5}g_1^2 M_1^2
-\frac 35 g_1^2 \xi_1+8\pi^2 \gamma m_L^2,\\
8\pi^2\frac{d m^2_E}{dt}&=&2\xi_\tau-\frac{12}{5}g_1^2 M_1^2+\frac 65 g_1^2\xi_1
+8\pi^2 \gamma m_E^2,\\
8\pi^2\frac{d m^2_{H_u}}{dt}&=&3\xi_t-3g_2^2 M_2^2-\frac{3}{5}g_1^2 M_1^2
+\frac 35 g_1^2\xi_1+8\pi^2 \gamma m_{H_u}^2,\\
8\pi^2\frac{d m^2_{H_d}}{dt}&=&3\xi_b+\xi_\tau-3 g_2^2 M_2^2-\frac{3}{5}g_1^2 M_1^2
-\frac 35 g_1^2 \xi_1+8\pi^2 \gamma m_{H_d}^2
\label{eqn:RGE_Hd},
\eea
\end{widetext}
where
\bea
\xi_t&=&y_t^2(m_{H_u}^2+m_Q^2+m_U^2+A_t^2),\\
\xi_b&=&y_b^2(m_{H_d}^2+m_Q^2+m_D^2+A_b^2),\\
\xi_\tau&=&y_\tau^2(m_{H_d}^2+m_L^2+m_E^2+A_\tau^2),
\eea
and
\bea
\xi_1&=&\half\left\{m_{H_u}^2 -m_{H_d}^2 \right. \nn \\
&+&\left. {\rm Tr}(m_Q^2-2m_U^2+m_D^2+m_E^2-m_L^2)\right\}.
\eea
The trace here means sum over the generations. 
Within our approximation only the $(3,3)$ family component of the three 
Yukawa matrices, $y_t, y_b, y_\tau$, are set to be nonzero.
The corresponding nonzero components of the three trilinear A-term 
matrices are $A_t, A_b, A_\tau$.

The above equations for the squarks and sleptons apply to the third family; the equations for the 
first and the second families are obtained from above by discarding the Yukawa and the A-term contributions (i.e. setting the $\xi_t$, $\xi_b$, $\xi_\tau$ terms to be zero).
All the other RG equations (the gauge couplings, 
the gaugino masses, the Yukawa couplings and the A-terms) 
are not affected by the hidden sector and are given e.g. in \cite{Castano:1993ri}.

\begin{figure}
\begin{center}
\includegraphics[scale=0.65]{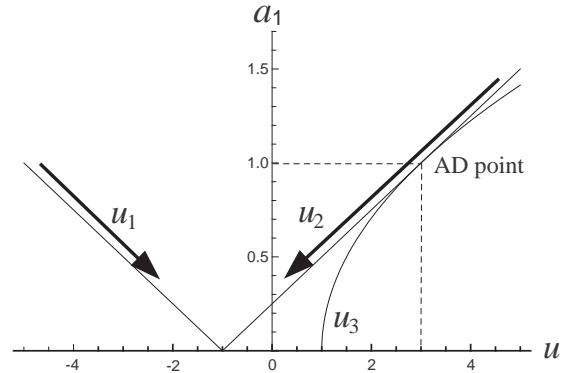}
\end{center}
\caption{\label{singular-f}
Flows of potential minima along the two dyon singular points.
}
\end{figure}

\subsection{The models of hidden-visible couplings}
\label{sec:models}
In the preceding section we described the ${\C N}=2$ supersymmetric $SU(2)\times U(1)$ gauge 
theory with $N_f=2$ matter hypermultiplets.
In order to use this as the hidden sector and discuss how it can affect the RG flow of the MSSM sector, one needs to identify the renormalization scale in the hidden sector.
A natural choice, which we shall adopt below, is to identify one of the hidden sector moduli as the renormalization scale $\mu$.
To describe, let us consider ${\cal N}=2$ SQED with one flavor for simplicity 
\cite{Seiberg:1994rs,Seiberg:1994aj}.
At generic points in the Coulomb branch the low-energy effective Lagrangian for the vector multiplet reads [cf. (\ref{eqn:VM})],
\begin{eqnarray} 
{\cal L}_{\rm VM}
=\frac{1}{8\pi}{\rm Im}\left[
\int d^4\theta A_D A^\dag
+{1 \over 2}\int d^2\theta {\partial A_D \over \partial A} 
W^{\alpha}W_\alpha
\right]. \label{eqn:qa}
\end{eqnarray}
The moduli space is parameterized by (the {\sc vev} of ) the scalar component of $A$.
It is of the special geometry type and the K\"{a}hler potential is identified with
\begin{equation}
K=\frac{1}{8\pi}{\rm Im} A_D A^\dag.
\end{equation}
The scalar kinetic term of (\ref{eqn:qa}) can be written using the K\"{a}hler metric
$K_{AA^\dag}\equiv\frac{\partial^2 K}{\partial A\partial A^\dag}$ as
\begin{equation}
K_{AA^\dag}AA^\dag.
\label{eqn:gkin}
\end{equation}
As the K\"{a}hler metric is renormalized, 
\begin{equation}
K_{AA^\dag}\rightarrow Z_A K_{AA^\dag},\label{eqn:Kren}
\end{equation}
from (\ref{eqn:gkin}) we see that the field $A$ undergoes
$A\rightarrow Z_A^{-1/2} A$.
When only the relative change matters, we may set $K_{AA^\dag}$ at some scale to be unity
and from ({\ref{eqn:Kren}) one may identify
$Z_A\equiv K_{AA^\dag}$.
We know that the quantum corrections to $K$ is one loop exact, leading to
($\Lambda$ being the Landau pole of QED), 
\begin{eqnarray}
A_D=-{i \over 2\pi}A\log{A \over \Lambda},
\end{eqnarray}
from which we may find the wave function renormalization coefficient,
\begin{equation}
Z_A\equiv \frac{\partial^2 K}{\partial A\partial A^\dag}=-\frac{1}{16\pi^2}\log\frac{A}{\Lambda}
+{\rm const.}
\end{equation}
This is the familiar solution to the RG equation, with $\mu$ replaced by $A$.
We thus see that the modulus $a\equiv \langle A\rangle$ plays the r\^{o}le of the 
renormalization scale $\mu$.

In our model there arises some ambiguity in choosing the renormalization scale, since the
moduli space is two-dimensional ($a_1$ and $u$).
The effective theory is derived at each point of the moduli $(a_1, u)$ and
it is natural to associate a renormalization group flow with the change of a modulus.
{\em We shall choose $a_1$ as the renormalization scale and identify it with $\mu$ in the 
RG analysis}, with appropriate rescaling.
%
This is also motivated by the fact that in our model the effective potential is lifted by the FI term 
in such a way that the {\sc vev} of the moduli rolls down toward the SUSY breaking 
vacuum at $(a_1, u)=(0, -\Lambda^2/8)$.
In this picture the RG flow is naturally realized as a flow from a nonzero $a_1$ to $a_1=0$.

The choice of a RG flow path in a multidimensional moduli space also involves some 
arbitrariness. 
We shall, following discussions of \cite{Gildener:1976ih,Sher:1988mj}, 
assume that a natural choice of a flow is along a trough of the effective potential.
In Seiberg-Witten type models such as ours, local minima of the potential typically occur at
singular points where light degrees of freedom (BPS states) condense. 
In our model there are three singular points in the hidden sector as shown in Fig.\ref{singular-t}.
Their positions in the $a_1$-$u$ moduli space are given by (\ref{eqn:s-f}).
Upon identifying $\mu=a_1$, the singular points move on the $a_1$-$u$ plane as the
renormalization scale changes.
Numerical study of the effective potential shows that in our model there are two troughs: 
one along the left dyon singularity and the other along the quark singularity 
connected to the right dyon singularity below the AD point.
We shall analyze RG flows along the singular points $u_1$ and $u_2$ from UV
(Fig.\ref{singular-f}), as $u_1$ and $u_2$ both flow to the SUSY breaking vacuum in 
the IR.
Along these flows, $u_{1}$ and $u_{2}$ are related to $a_1$ by (\ref{eqn:s-f}) and renormalized 
quantities such as the gauge coupling constant are all functions of $\mu=a_1$.

The messenger fields $\Psi$ and $\tilde\Psi$ charged under the MSSM gauge groups may
couple to the hidden sector fields in various ways. We consider following two cases:

\begin{description}
  \item[(1) Coupling to $A_1$]
  
  -- We choose the superpotential of the messenger fields $\Psi$, $\tilde\Psi$ as
  \beq
  W_{\rm mess}=A_1 \Psi\tilde\Psi +M_{\rm m}\Psi\tilde\Psi. \label{model1}
  \eeq
  Here, $M_{\rm m}$ is the messenger mass which is taken to be $10^{13}$ GeV in 
  the numerics.
  The hidden sector field $X$ is $A_1$ and its boundary condition at the SUSY breaking vacuum
  is $\langle X\rangle=\langle A_1\rangle =0$.
  The breaking scale is $\langle F_1\rangle \sim \lambda$.

  \item[(2) Coupling to $A_2$]
  -- The messenger superpotential is
  \beq
  W_{\rm mess}=\frac{u}{M_{\rm c}}\Psi\tilde\Psi+M_{\rm m}\Psi\tilde\Psi \label{model2},
  \eeq
  where $M_{\rm m}$ is the messenger mass as above, and $u=\Tr (A_2^2)$.
  The mass scale $M_{\rm c}$ has been introduced since   
  $u$ is a higher dimensional operator.
  The hidden sector field is $X=u/M_{\rm c}$.
  The boundary condition at the SUSY breaking vacuum is $\langle u\rangle =-\Lambda^2/8$.
  The breaking scale is $\langle F_2\rangle \sim \lambda$.
\end{description}

Note that $\langle X\rangle$ is much smaller than $M_{\rm m}$ and does not affect the GMSB
boundary conditions (\ref{eqn:GauginoMass}), (\ref{smass}).
The hidden sector field $X$ undergoes wave function renormalization as
$X\rightarrow Z_X^{-n_X/2} X$,
where $n_X$ is the scaling dimension of the operator coupled to the messenger field
($n_X=1$ for $X=A_1$ and $n_X=2$ for $X=u/M_{\rm c}$).
The anomalous dimension is written using the coefficient $Z_X$ as
\beq
\gamma(t)=-n_X\mu\frac{d}{d\mu}\ln Z_X
=-n_X\frac{d}{dt}\ln Z_X. \label{a-d}
\eeq
Using this in (\ref{eqn:RGE-s1}) we obtain,
\begin{eqnarray}
k_i(t)&=&
k_i(0)\left(\frac{Z_X(0)}{Z_X(t)}\right)^{n_X} \nn\\
&&\hspace{-10mm}+\frac{1}{16\pi^2}\sum_{a=1}^3 8 C_2^a(R_i)\!\!\int_t^0\!\! ds g_a^6(s) 
\left(\frac{Z_X(s)}{Z_X(t)}\right)^{n_X}\!\! G_a. 
\label{sol-r}
\end{eqnarray}
As discussed in \cite{Giudice:1997ni,ArkaniHamed:1998kj}, the effects of soft SUSY breaking 
are all included in the wave function renormalization,
and once the latter is known there is no need to compute separate Feynman diagrams.

The wave function renormalization of the hidden sector fields is induced by change of the 
K\"{a}hler metric, 
$K_{A_iA_i^\dag}\equiv \partial^2 K/\partial A_i\partial A_i^\dag
\rightarrow Z_{A_i} K_{A_iA_i^\dag}$, and just as in the SQED case above,
one may set $K_{A_iA_i^\dag}$ at some scale to be unity since only the relative change matters
in the RG study.
Thus the renormalization coefficient $Z_{A_i}$ is identified with $K_{A_iA_i^\dag}$, and hence with $b_{ii}$ of (\ref{eqn:bij}).
Concretely, in the $A_1$-coupled case, 
\beq
Z_{A_1}\equiv\frac{\partial^2 K}{\partial A_1\partial A_1^\dag}=
\frac{1}{8\pi}{\rm Im}\frac{\partial^2}{\partial A_1\partial A_1^\dag} 
\Big(\frac{\partial{\C F}}{\partial A_1}A_1^\dag\Big)
=b_{11}. \label{w-r}
\eeq
Similarly, in the $A_2$-coupled case we have $Z_u\equiv Z_{A_2}=b_{22}$.
While the semiclassical picture is only applicable in the UV regions, we can follow the
RG flow down to IR since $b_{11}$ and $b_{22}$ are globally solved using the exact method.

%
%
\begin{figure*}
\begin{eqnarray*}
\begin{array}{ccc}
\includegraphics[scale=0.65]{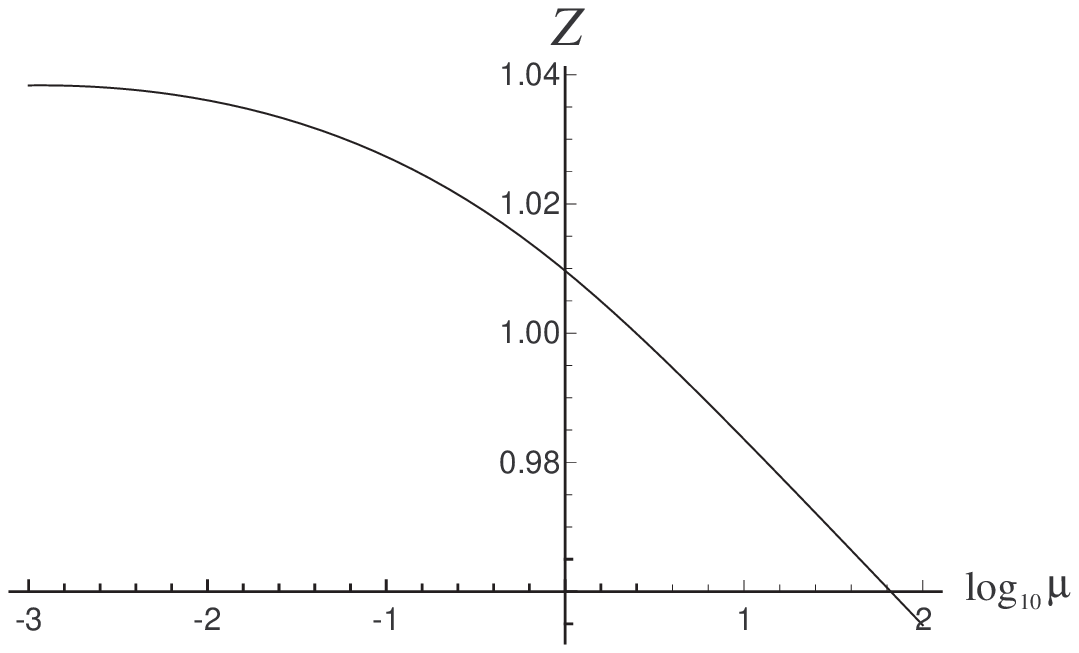}
&&
\includegraphics[scale=0.65]{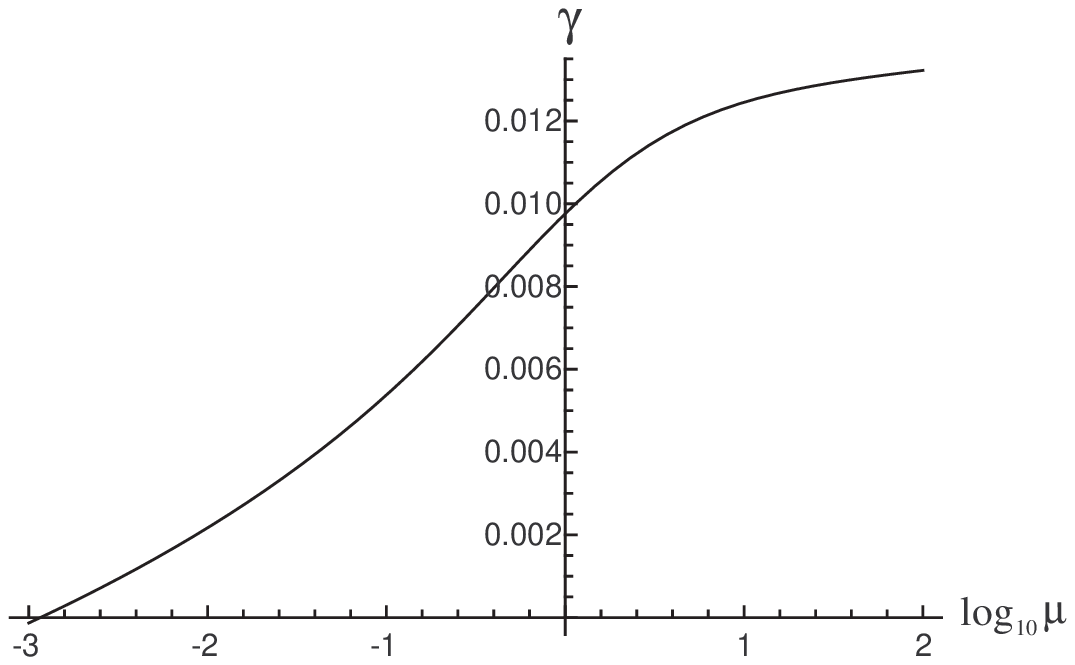}
\end{array}
\end{eqnarray*}
\caption{\label{Z-L1}The wave function renormalization coefficient $Z=b_{11}$ (left) and the anomalous dimension $\gamma$ (right) along the flow $u_1$ in the $A_1$-coupled model.}
\end{figure*}
\begin{figure*}
\begin{eqnarray*}
\begin{array}{ccc}
\includegraphics[scale=.7]{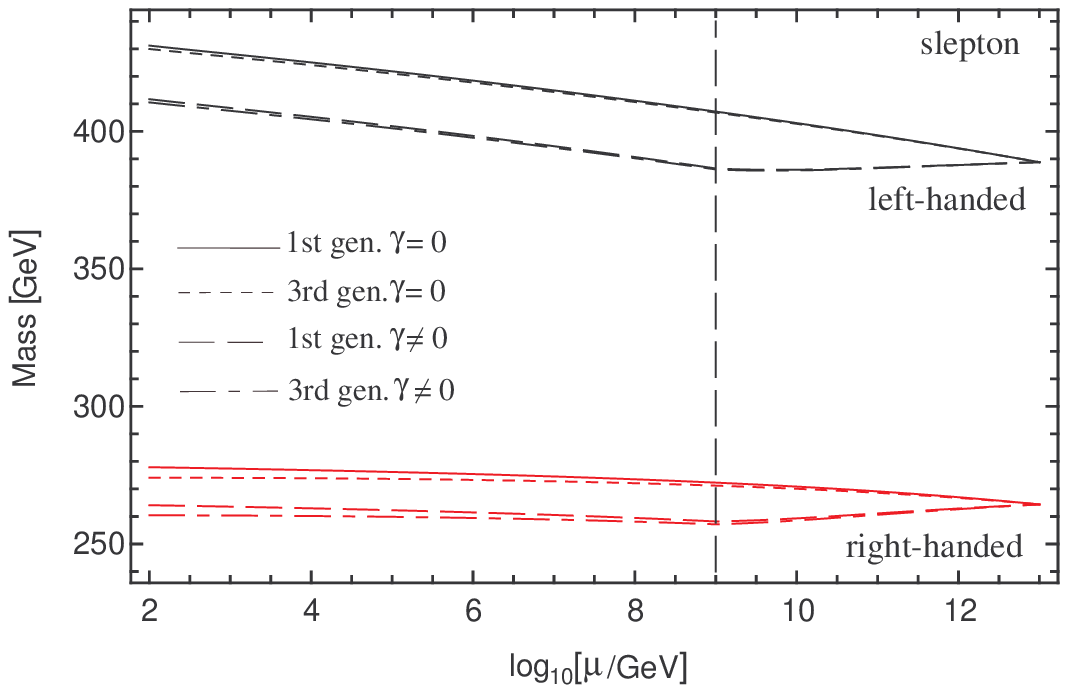}
&&
\includegraphics[scale=.7]{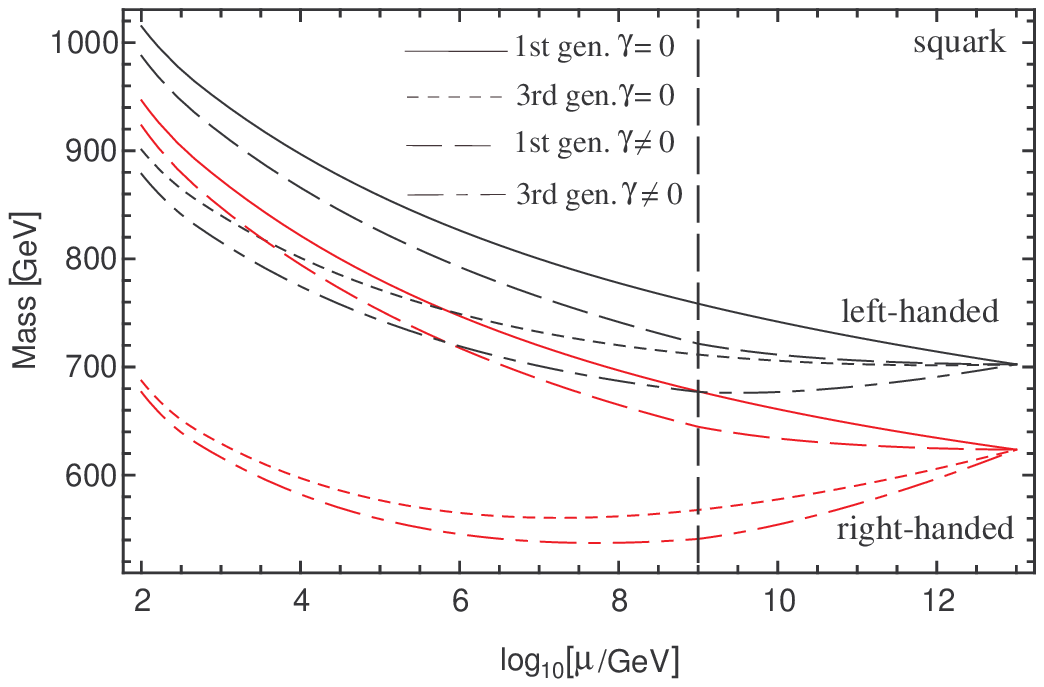}
\end{array}
\end{eqnarray*}
\caption{\label{sleptonL}
The mass RG flows for the first and third generation sleptons (left panel) and squarks 
(right panel), with ($\gamma\neq 0$) and without ($\gamma=0$) hidden sector effects. 
Here, the messenger is coupled to $A_1$, and $\gamma$ is the anomalous dimension for $A_1$ 
shown in Fig.\ref{Z-L1}.
The flow is taken along the $u_1$ (left dyon) singularity.
We have rescaled $\mu \rightarrow (10^{9}/\tilde{M}_{\rm hid}) \mu$, where 
$\tilde{M}_{\rm hid}=0.2$ is the hidden scale before rescaling. 
We have chosen $M_{\rm hid}=10^9$ GeV (the vertical dashed line), 
$M_{\rm m} = 10^{13}$ GeV, and $\tan\beta = 10$.}
\end{figure*}

\section{Numerical results}

In this section we analyze the RG flows of the MSSM scalar masses for the models (\ref{model1}) and (\ref{model2}).
We solve the RG equations (\ref{eqn:RGE_Q})-(\ref{eqn:RGE_Hd}) numerically along the two 
possible flows $u_1$ and $u_2$ of Fig.\ref{singular-f}. 
The coupling to the hidden sector affects the MSSM RG flows between the messenger scale 
$\mu=M_{\rm m}$ and the hidden sector scale $\mu=M_{\rm hid}$, through the anomalous 
dimension $\gamma(t)$.
We first evaluate the anomalous dimension (\ref{a-d}) and the wave function renormalization 
scale (\ref{w-r}).
We follow the conventions of Sec. \ref{sec:effpot} and set the dynamical scale to be
$\Lambda=2\sqrt{2}$, and then rescale it in order to suit the minimal GMSB scenario with 
high messenger scale.

In our numerical calculations we made the following approximations that are standard in 
the GMSB RG analysis:
only the $(3,3)$ family component of the three Yukawa matrices, and only the $(3,3)$ family component of the trilinear A-term matrices, are set to be nonzero. 
The latter are assumed to vanish at the messenger scale. 
We use the following values: 
$\tan\beta=10$, the soft mass scale $M_S=500$ GeV, and the messenger scale $M_{\rm m}=10^{13}$ GeV. 
The hidden sector scale is $M_{\rm hid}=\langle F_X\rangle^{1/2}=10^9$ GeV, above which the
hidden sector degrees of freedom are integrated out.

\subsection{$A_1$-coupled messenger}

\subsubsection{Flow along the left dyon singularity}\label{A1L}
Along the flow $u_1$, the wave function renormalization scale and the anomalous 
dimension are
\begin{eqnarray}
Z_{A_1} &=& b_{11}(u_1(a_1), a_1){\Big |}_{a_1 = \mu}, \label{Z-L1A} \\
\gamma_{A_1} &=& - a_1\frac{d}{da_1}Z_{A_1}(u_1(a_1), a_1){\Big |}_{a_1 = \mu}. 
\label{gamma-L1A}
\end{eqnarray}
In Fig.\ref{Z-L1} we show numerical plots of (\ref{Z-L1A}) and (\ref{gamma-L1A}) along the
renormalization scale $\mu$, with $\Lambda=2\sqrt{2}$. 
In the phenomenological setting we rescale 
$\mu\rightarrow (10^9/\tilde{M}_{\rm hid})\mu$,
where $\tilde{M}_{\rm hid}=0.2$ is the hidden scale before rescaling\footnote{
For $\tilde{M}_{\rm hid}\lesssim 0.2$ the condensations (\ref{cond}) around the left and 
right dyon points overlap, signaling the breakdown of the local low-energy effective theory 
description. 
For this reason we shall adopt $\tilde{M}_{\rm hid}=0.2$ when analyzing the left 
dyon singularity, also in the $A_2$-coupled model below.
}.
%
%
The RG equations (\ref{eqn:RGE_Q})-(\ref{eqn:RGE_Hd}) are then solved numerically, using
the anomalous dimension (\ref{gamma-L1A}) with the rescaled $\mu$.
The results are shown in Fig.\ref{sleptonL} for the slepton and squark masses, with and without 
the hidden sector effects.
For both sleptons and quarks, the hidden sector effects are seen to decrease the masses, 
as the scale goes down from the messenger to the hidden scale.
This behavior is understood by looking at Eq. (\ref{sol-r}). 
The effects of the hidden sector come in $Z_X(0)/Z_X(t)$ and $Z_X(s)/Z_X(t)$. 
The contribution from the latter [the second term of (\ref{sol-r})] is suppressed by the gauge 
coupling constants, and thus the dominant contribution is from the former (the first term).
We see from Fig.\ref{Z-L1} that $Z_{A_1}(0)/Z_{A_1}(t)<1$,
leading to the smaller masses of Fig.\ref{sleptonL}.

%
%
\begin{figure*}
\begin{eqnarray*}
\begin{array}{ccc}
\includegraphics[scale=0.65]{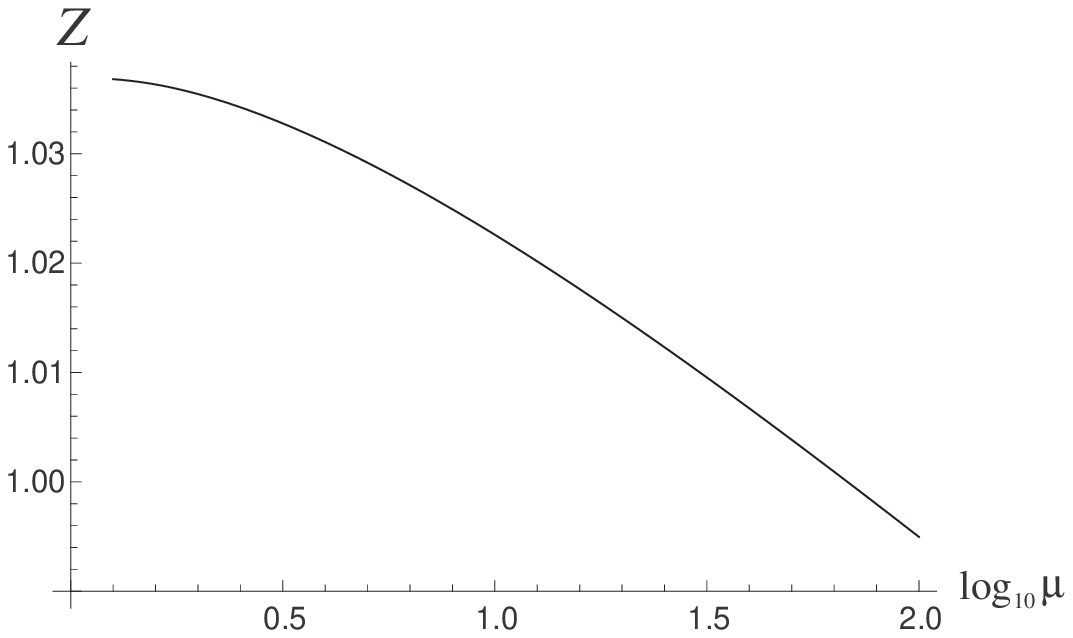}
&&
\includegraphics[scale=0.65]{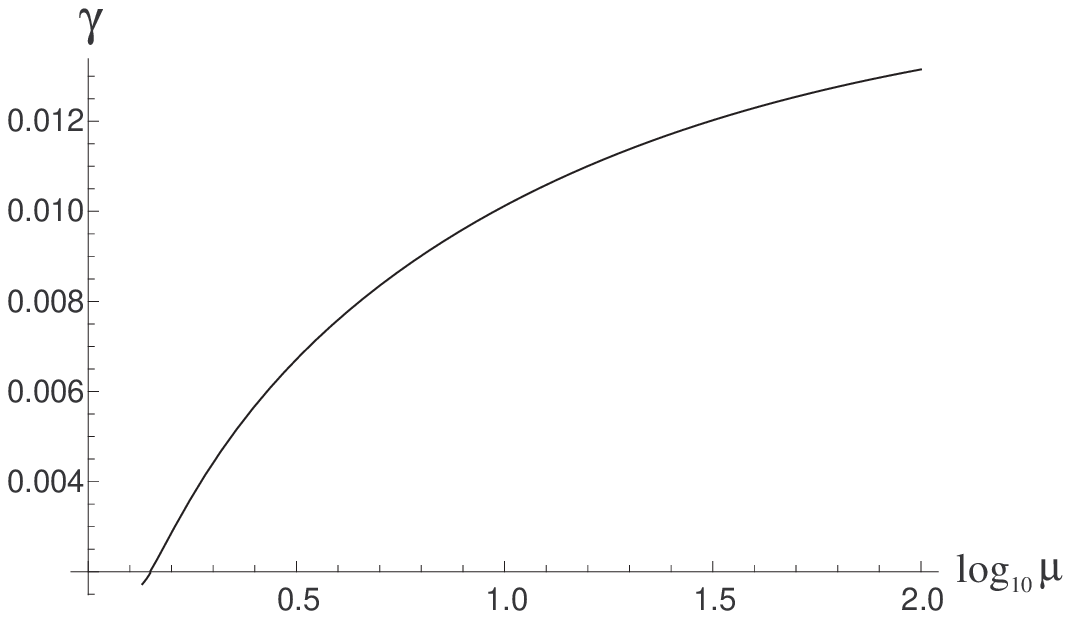}
\end{array}
\end{eqnarray*}
\caption{\label{Z-R1}The wave function renormalization coefficient $Z=b_{11}$ (left)
and the anomalous dimension $\gamma$ (right) along the flow $u_2$ in the 
$A_1$-coupled model.}
\end{figure*}
%
\begin{figure*}
\begin{eqnarray*}
\begin{array}{ccc}
\includegraphics[scale=.7]{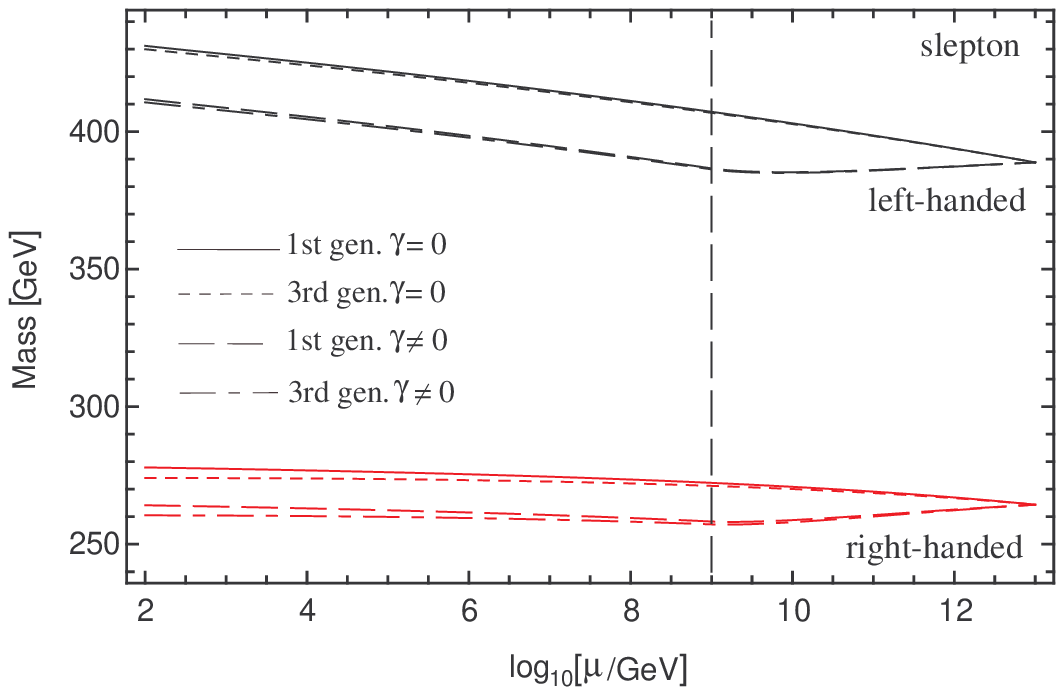}
&&
\includegraphics[scale=.7]{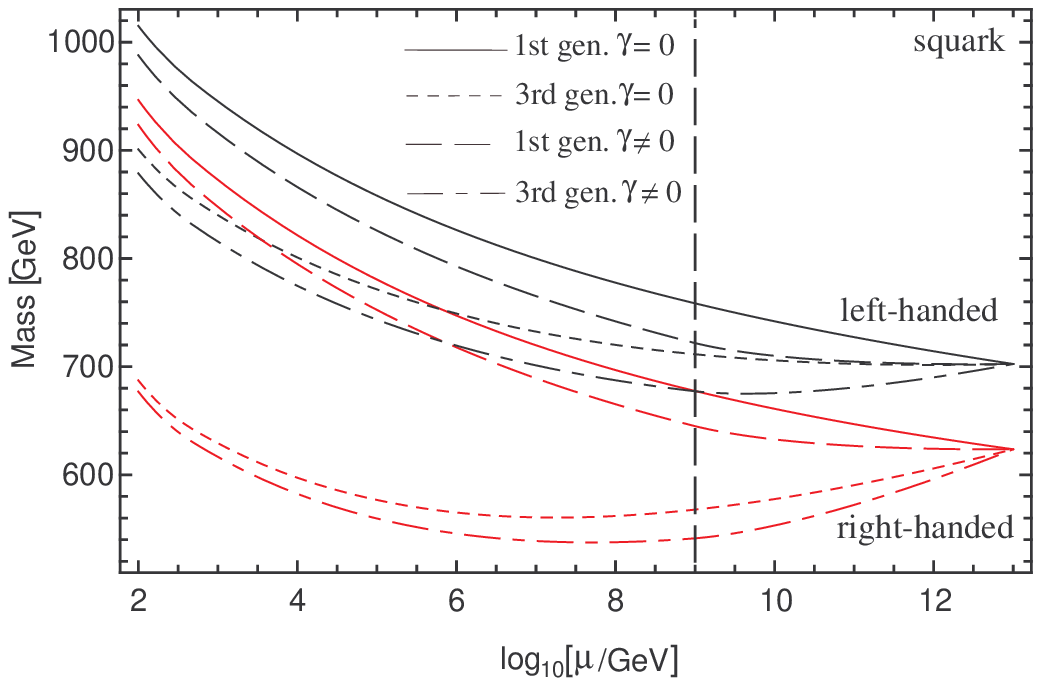}
\end{array}
\end{eqnarray*}
\caption{\label{slqR-s}
The mass RG flows for the first and third generation sleptons (left panel) and squarks 
(right panel), with ($\gamma\neq 0$) and without ($\gamma=0$) hidden sector effects. 
The vertical dashed line indicates the hidden scale $M_{\rm hid}=10^9$ GeV.
The messenger is coupled to $A_1$ and the flow is taken along the $u_2$ (right dyon) singularity.
We have rescaled 
$\mu \rightarrow (10^{9}/\tilde{M}_{\rm hid}) \mu$, where $\tilde{M}_{\rm hid}=1.2$. 
We have chosen $M_{\rm m} = 10^{13}$ GeV and $\tan\beta = 10$.}
\end{figure*}

\subsubsection{Flow along the right dyon singularity}
We next consider the RG flows of the MSSM scalar masses along $u_2$. 
In this case, the flow is somewhat complicated as it contains the AD point. 
At the AD point ($\mu=a_1=1$ in our scale $\Lambda=2\sqrt{2}$)
the theory becomes superconformal and the local effective theory description 
breaks down. 
In our analysis we set the hidden scale to be $\tilde{M}_{\rm hid}=1.2$, i.e.
above the AD point so that the hidden sector dynamics at the AD point does not contribute to 
the RG flow.
In this case the wave function renormalization coefficient and the anomalous dimension are
\begin{eqnarray}
Z_{A_1} &=& b_{11}(u_2(a_1), a_1){\Big |}_{a_1 = \mu}, \label{Z-R1A} \\
\gamma_{A_1} &=& - a_1\frac{d}{da_1}Z_{A_1}(u_2(a_1), a_1){\Big |}_{a_1 = \mu}.
\label{gamma-R1A}
\end{eqnarray}
These are shown in Fig.\ref{Z-R1}. 
Numerical solutions to the RG equations (\ref{eqn:RGE_Q})-(\ref{eqn:RGE_Hd}),
with rescaling $\mu \rightarrow (10^9/\tilde{M}_{\rm hid}) \mu$, are shown in 
Fig.\ref{slqR-s}. 
We see from the messenger scale down to the hidden scale that the hidden sector effects 
render both slepton and squark masses smaller. 
Like in the $u_1$ case, this is due to $Z_{A_1}(0)/Z_{A_1}(t)<1$ in Eq. (\ref{sol-r}).

\subsection{$A_2$-coupled messenger}

\begin{figure*}
\begin{eqnarray*}
\begin{array}{ccc}
\includegraphics[scale=0.7]{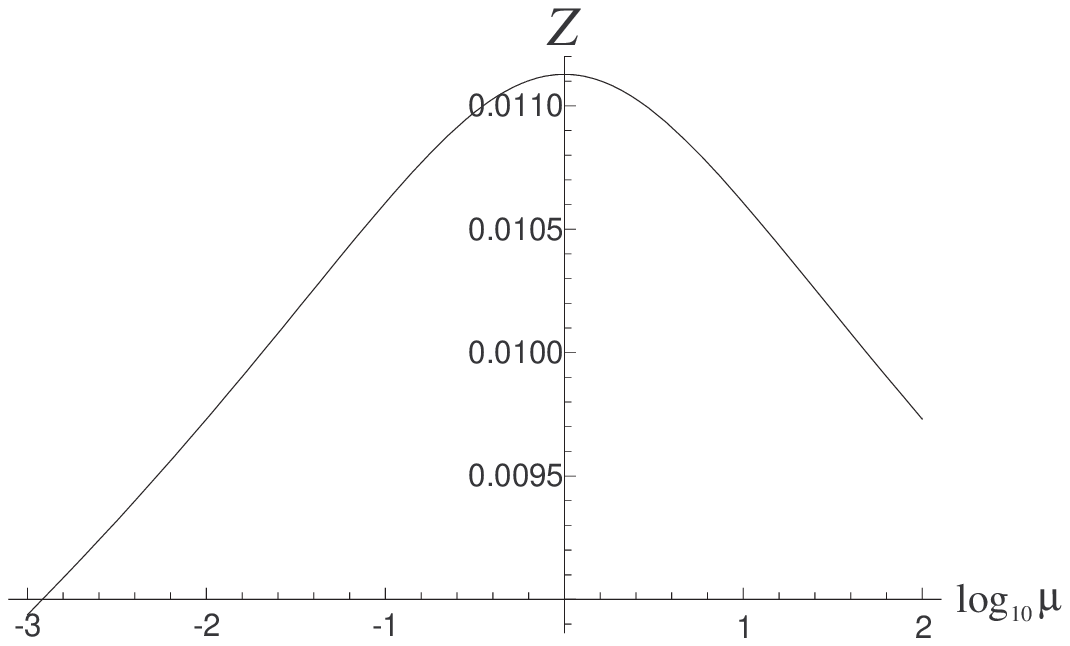}
&&
\includegraphics[scale=0.7]{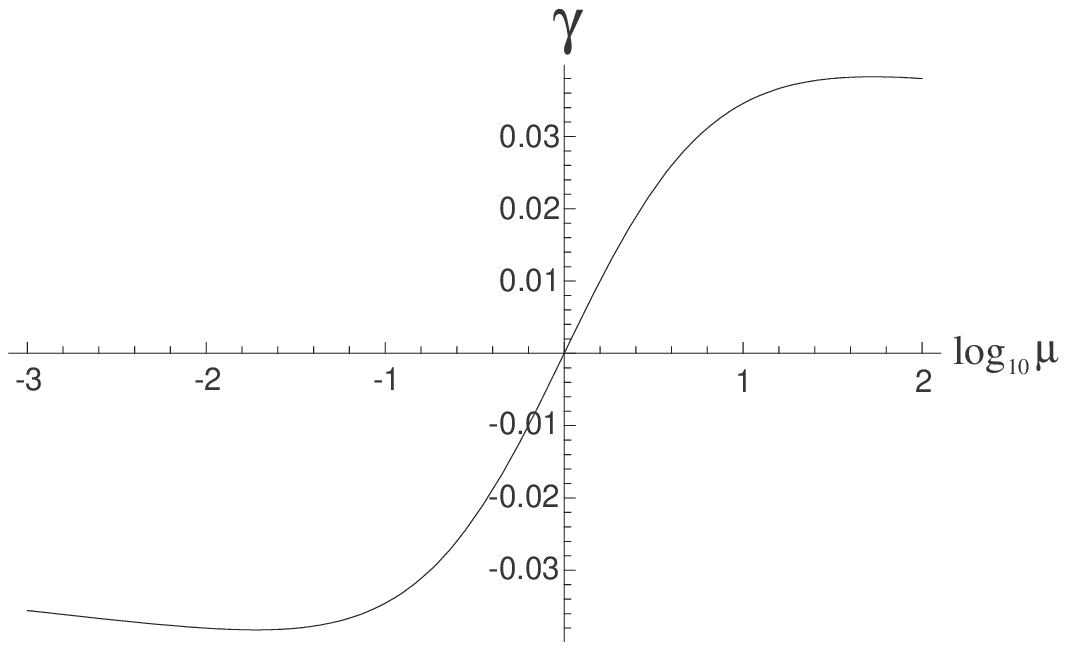}
\end{array}
\end{eqnarray*}
\caption{\label{Z-L2}
The wave function renormalization coefficient $Z=b_{22}$ (left) and the anomalous dimension 
$\gamma$ (right) along the flow $u_1$ in the $A_2$-coupled model.}
\end{figure*}
\begin{figure*}
\begin{eqnarray*}
\begin{array}{ccc}
\includegraphics[scale=.7]{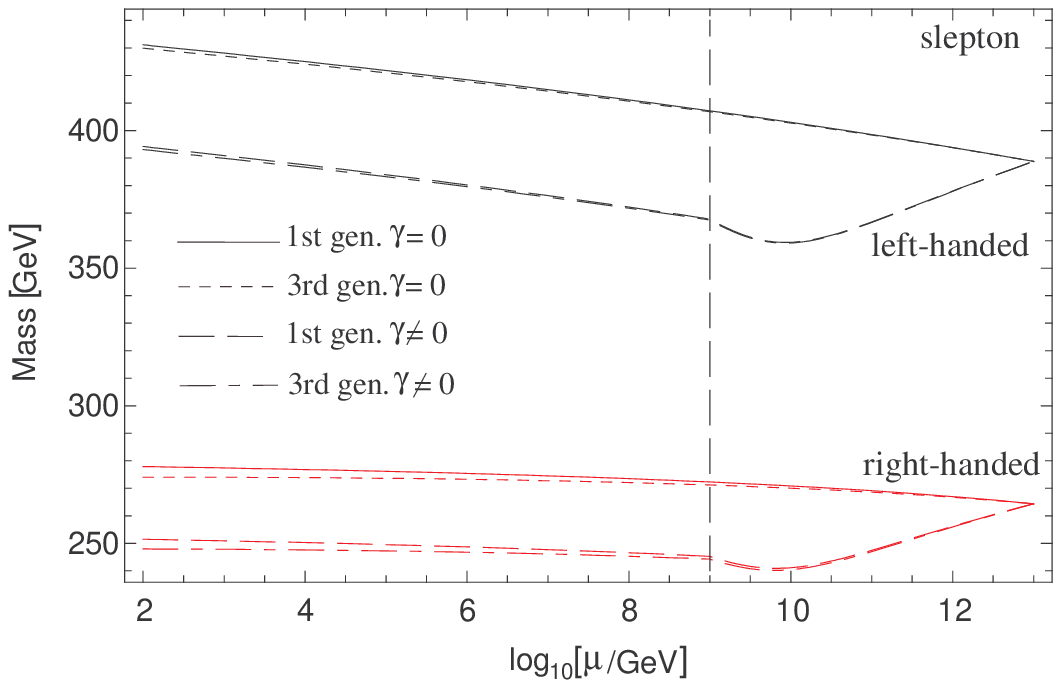}
&&
\includegraphics[scale=.7]{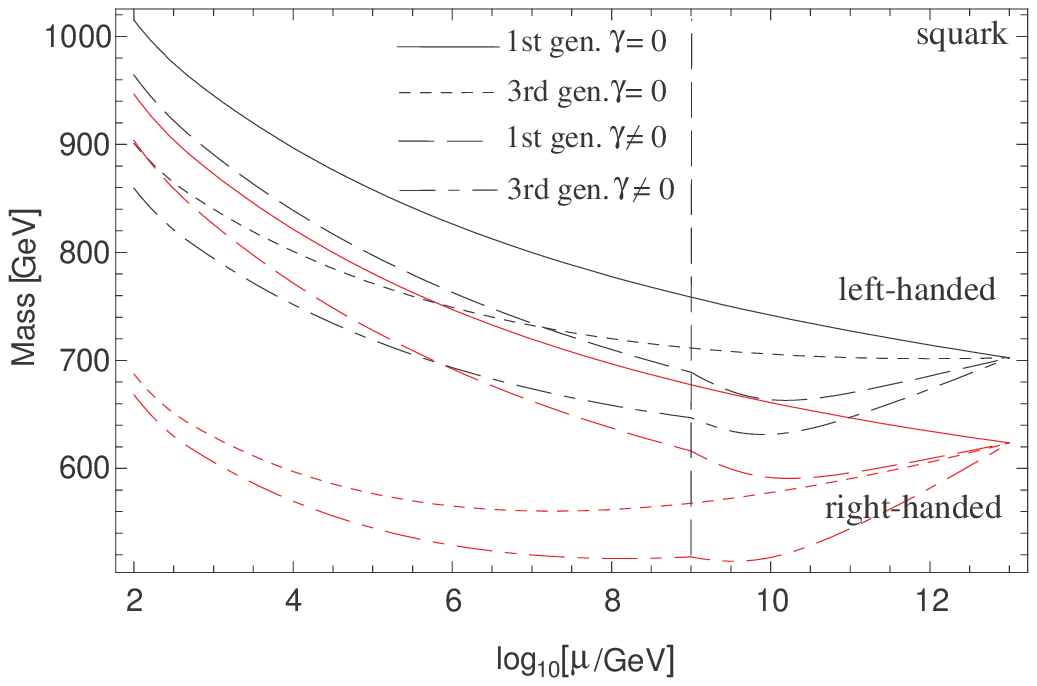}
\end{array}
\end{eqnarray*}
\caption{\label{slqL-2}
The mass RG flows for the first and third generation sleptons (left panel) and squarks 
(right panel), with ($\gamma\neq 0$) and without ($\gamma=0$) hidden sector effects. 
The vertical dashed line indicates the hidden scale $M_{\rm hid}=10^9$ GeV.
The messenger is coupled to $A_2$ and the flow is taken along the $u_1$ (left dyon) singularity.
We have rescaled 
$\mu \rightarrow (10^{9}/\tilde{M}_{\rm hid}) \mu$, with the hidden scale before rescaling 
$\tilde{M}_{\rm hid}=0.2$. 
We have chosen $M_{\rm m} = 10^{13}$ GeV and $\tan\beta = 10$.}
\end{figure*}

\subsubsection{Flow along the left dyon singularity}
Let us discuss the $A_2$-coupled mass RG flow (\ref{model2}).
We first consider the flow along $u_1$. 
In this model, the wave function renormalization coefficient and the anomalous dimension are
\begin{eqnarray}
Z_{A_2} &=& b_{22}(u_1(a_1), a_1){\Big |}_{a_1=\mu}, \\
\gamma_{A_2} &=& -2a_1\frac{d}{d a_1}Z_{A_2}(u_1(a_1), a_1){\Big |}_{a_1=\mu}.
\end{eqnarray}
These are shown in Fig.\ref{Z-L2}. 
In contrast to the previous examples we see that $Z_{A_2}$ increases
as the scale goes up to $\mu=1$ and monotonically decreases above that scale. 
Accordingly, $\gamma_{A_2}$ changes its sign at $\mu = 1$. 
Fig.\ref{slqL-2} shows the mass RG flows of the sleptons and the squarks, obtained by
setting $\mu \rightarrow (10^{9}/\tilde{M}_{\rm hid}) \mu$ and
$\tilde{M}_{\rm hid}=0.2$.
In this setting, we see that the hidden sector gives rise to different effects on the MSSM
mass spectrum for $\mu>M'$ and $\mu<M'$, where $M'=5\times 10^9$ GeV is the energy scale 
corresponding to $\mu=1$ before rescaling.
When $M'<\mu<M_{\rm m}$, the effect of the hidden sector is to decrease the scalar masses, 
which is understood from (\ref{sol-r}) with $Z_{A_2}(0)/Z_{A_2}(t_{M'})<1$ 
[where $t_\mu\equiv\ln(\mu/M_{\rm m})$], the same effect as in the $A_1$-coupled model.
For $\tilde{M}_{\rm hid}<\mu<M'$ the effect of the hidden sector increases the soft scalar masses,
resulting from $Z(t_{M'})/Z(t_{M_{\rm hid}})>1$.
In fact, the scale $M'$ is considered to be a boundary between the weak and strong coupling 
regions.
The messenger fields are coupled to $u/M_{\rm c}$, where $u=\Tr(A_2^2)$ semiclassically
and $A_2$ is a good variable in the weak coupling region.
We have extrapolated $Z_{A_2}=b_{22}$ down to the strong coupling region using the exact
solution.   
Below the scale $M'$ the nature of the coupling changes, leading to the increasing effect of the
MSSM scalar masses.

\begin{figure*}
\begin{eqnarray*}
\begin{array}{ccc}
\includegraphics[scale=0.7]{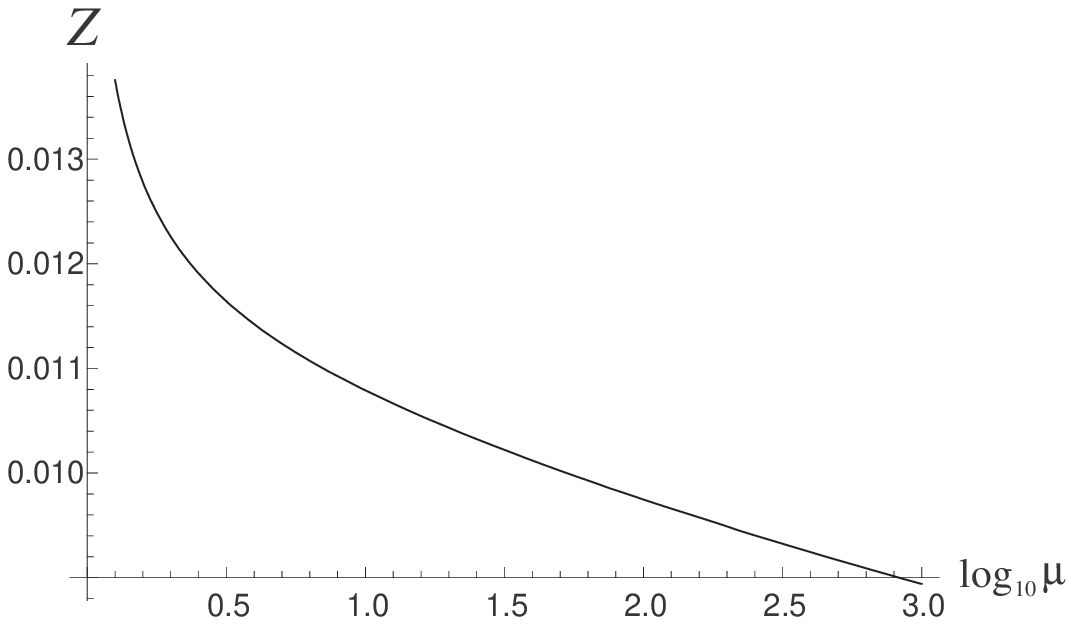}
&&
\includegraphics[scale=0.7]{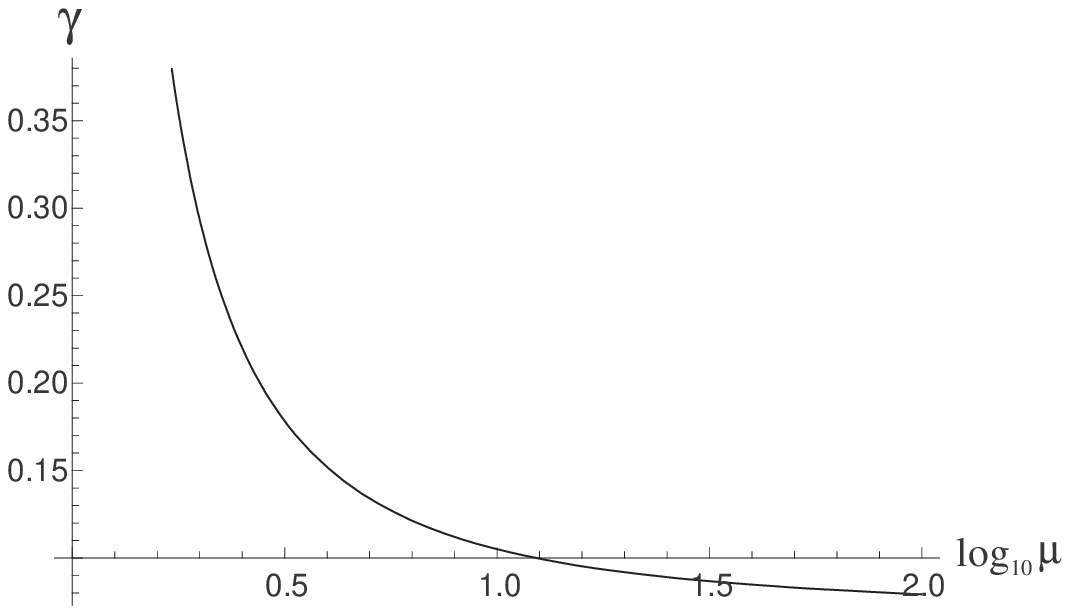}
\end{array}
\end{eqnarray*}
\caption{\label{Z-2R}
The wave function renormalization coefficient $Z=b_{22}$ (left) and the anomalous dimension 
$\gamma$ (right) along the flow $u_2$ in the $A_2$-coupled model.}
\end{figure*}
\begin{figure*}
\begin{eqnarray*}
\begin{array}{ccc}
\includegraphics[scale=.7]{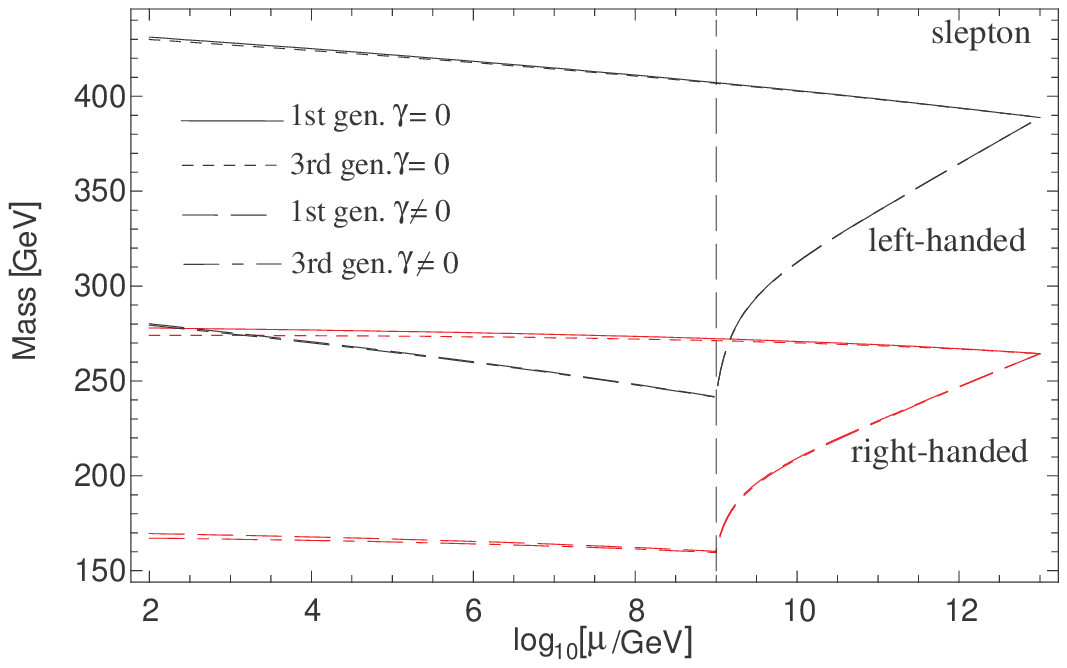}
&&
\includegraphics[scale=.7]{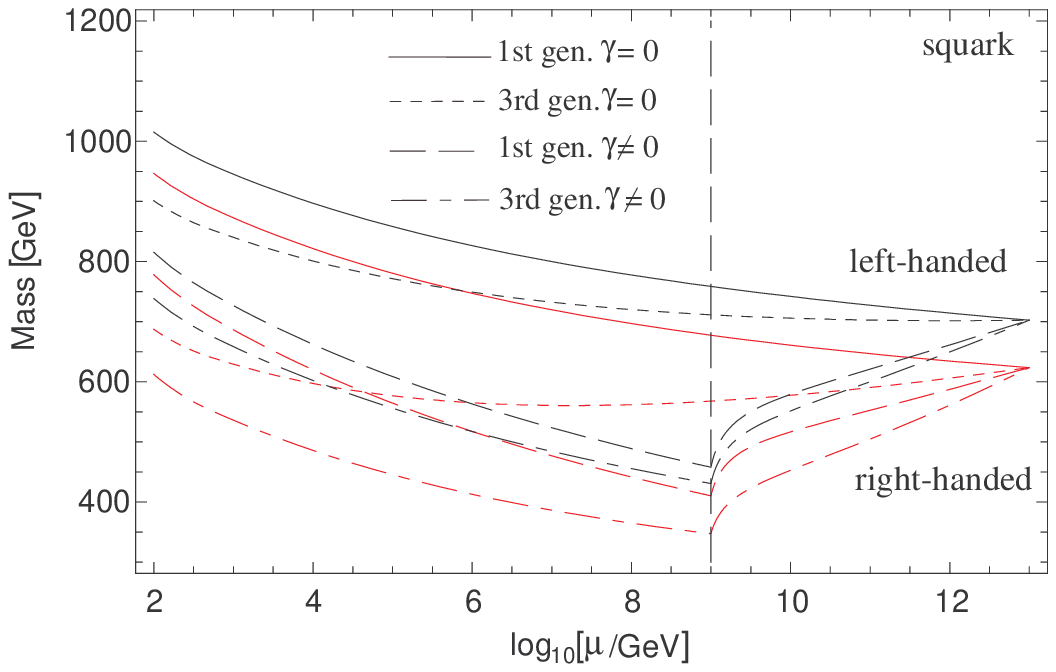}
\end{array}
\end{eqnarray*}
\caption{\label{slqL-A2R}
The mass RG flows for the first and third generation sleptons (left panel) and squarks 
(right panel), with ($\gamma\neq 0$) and without ($\gamma=0$) hidden sector effects. 
The vertical dashed line indicates the hidden scale $M_{\rm hid}=10^9$ GeV.
The messenger is coupled to $A_2$ and the flow is taken along the $u_2$ (right dyon) singularity.
We have rescaled 
$\mu \rightarrow (10^{9}/\tilde{M}_{\rm hid}) \mu$, with the hidden scale before rescaling 
$\tilde{M}_{\rm hid}=1.2$. 
We have chosen $M_{\rm m} = 10^{13}$ GeV and $\tan\beta = 10$.}
\end{figure*}

\subsubsection{Flow along the right dyon singularity}

Finally we consider the $A_2$-coupled model along the flow $u_2$. 
As in the $A_1$-coupled case along $u_2$, we take the hidden scale 
to be $\tilde{M}_{\rm hid}=1.2$ so that the dynamics around the AD point $\mu=1$
does not contribute to the mass RG flow.
The wave function renormalization coefficient and the anomalous dimension,
\begin{eqnarray}
Z_{A_2} &=& b_{22}(u_2(a_1), a_1){\Big |}_{a_1=\mu}, \\
\gamma_{A_2} &=& -2a_1\frac{d}{d a_1}Z_{A_2}(u_2(a_1), a_1){\Big |}_{a_1=\mu},
\end{eqnarray}
above $\tilde{M}_{\rm hid}=1.2$ are shown in Fig.\ref{Z-2R}.
In this case, the anomalous dimension rapidly increases as the scale approaches
$\tilde{M}_{\rm hid}$ from a higher scale.
Numerical solutions to the RG equations (\ref{eqn:RGE_Q})-(\ref{eqn:RGE_Hd}),
with rescaling $\mu\rightarrow (10^9/\tilde{M}_{\rm hid})\mu$, are shown in Fig.\ref{slqL-A2R}. 
We see that both slepton and squark masses are decreased as the scale goes down 
from $M_{\rm m}$ to $\tilde{M}_{\rm hid}$.
We observe sharp decline of the mass RG flow near $\tilde{M}_{\rm hid}$, which is
expected from the behavior of the anomalous dimension.

In the numerical study we chose $\tilde{M}_{\rm hid}=1.2$ in order to avoid the AD point
where the local effective theory description becomes unreliable.
If we nevertheless take $\tilde{M}_{\rm hid}$ close to the AD point and take the results at face 
value, we have an extremely large anomalous dimension and a very steep decline of the mass 
RG flow. 
If this is indeed the case it is phenomenologically very interesting, 
since the lightest scalar tau (stau) can easily be lighter than the bino, leading to
rich implications in collider physics \cite{Arai:2010ds}.
Furthermore, with the large anomalous dimension, one may possibly solve the $\mu$ problem 
of the GMSB scenario \cite{Roy:2007nz,Murayama:2007ge}.
While a reliable effective theory at the AD point is lacking at the moment, 
it would certainly be worthwhile investigating these possibilities further.

In the present setting with $\tilde M_{\rm hid}=1.2$ and $\tan\beta=10$, we have the stau mass
$m_{\tilde{\tau}}=167.8$ GeV and the bino mass $M_1=134.9$ GeV at the low energy. 
We find that the stau can be lighter than the bino with moderate values of $\tan\beta$: 
with $\tan\beta=25$, for example, we have stau mass $m_{\tilde{\tau}}=132.9$ 
GeV and bino mass $M_1=134.9$ GeV.

\section{Summary and Discussions}
\label{sec:concl}


In this paper we have discussed how a strongly coupled hidden sector dynamics may affect
the MSSM low-energy mass spectrum,
using a calculable model of hidden sector and the minimal GMSB scenario as a concrete
example.
We found that the effects on the RG flow can be qualitatively different from the perturbative toy 
model cases discussed in \cite{Campbell:2008tt,Arai:2010ds}.
In our strongly coupled example the hidden sector effects can make the scalar masses
larger or smaller, depending sensitively on the coupling of the messenger fields to the hidden
sector fields, as well as on the dynamics of the hidden sector itself; 
this is in a stark contrast to the perturbative case where the scalar masses can 
only be smaller.
In fact, this feature is not quite surprising since non-Abelian gauge fields have different RG
behavior than scalar fields or Abelian gauge fields.
We expect our finding to be generic in strongly coupled hidden sector models.


Let us touch upon some phenomenological implications of the $A_1$- and 
$A_2$-coupled examples given in Sec.\ref{sec:models}.
These two models are actually quite different. 
In the $A_1$-coupled model (\ref{model1}) the gravitino mass is
\beq
m_{3/2}\sim\frac{\langle F_{A_1}\rangle}{\sqrt 3 M_{\rm Pl}},
\eeq
which is $m_{3/2}\sim 1$ GeV with our parameter values
($M_{\rm Pl}=2.4\times 10^{18}$ GeV is the reduced Planck mass).
The gravitino of this mass is the lightest superparticle (LSP) and it is a good candidate of 
cold dark matter.
For a typical neutralino mass range of a few hundred GeV, this gravitino mass is marginally 
compatible with the big-bang nucleosynthesis (BBN) bound \cite{Kawasaki:2008qe}.
In the $A_2$-coupled model (\ref{model2}) on the other hand, the gravitino mass is evaluated as
\beq
m_{3/2}\sim\frac{\langle F_{A_2}\rangle}{\sqrt 3 M_{\rm Pl}}
=\frac{M_{\rm c}}{\Lambda}\frac{\langle F_{X}\rangle}{\sqrt 3 M_{\rm Pl}}
\sim\frac{M_{\rm c}}{\Lambda}\times (1\mbox{ GeV}),
\eeq
where $\sqrt{\langle F_X\rangle}\sim 10^9$ GeV is assumed.
If $M_{\rm c}/\Lambda\gtrsim {\cal O}(100)$, the gravitino cannot be
the LSP anymore and in this case, the usual neutralino LSP scenario
is applicable. For the parameter set in Fig.10, we have $m_{3/2} \sim 1$ TeV.

Finally, we comment that the RG analysis presented in this paper is also possible for other 
models of strongly coupled hidden sector, as long as they are of perturbed 
Seiberg-Witten type and the K\"ahler metric is calculable.
Indeed, there are many models of this type, e.g. \cite{Ooguri:2007iu,Marsano:2007mt},
studied in the context of metastability of supersymmetric vacua \cite{Intriligator:2006dd}.
The hidden sector RG provides a test bed for studying dynamics of
the hidden sector and the mechanism of SUSY breaking on observational basis,
and it is certainly very interesting to discuss phenomenological implications of various 
hidden sector models.


\hspace{10mm}

\subsection*{Acknowledgments}
M.~A. and S.~K. thank the organizers of the YITP workshop YITP-W-08-04 on 
``Development of Quantum Field Theory and String Theory.''
This work was supported in part by the Research Program MSM6840770029,
the Project of International Cooperation ATLAS-CERN of the Ministry of Education, 
Youth and Sports of the Czech Republic (M.~A.), 
the WCU grant R32-2008-000-10130-0 (S.~K.), and 
by the DOE Grant \#DE-FG02-10ER41714 (N.~O.).

\bigskip
\begin{appendix}

\section{${\C N}=2$ technicalities}

In this Appendix we give technical details of the ${\C N}=2$ supersymmetric QCD
that are used in actual computations of the hidden sector \cite{Arai:2001pi,Arai:2007md}.
A basic assumption on our hidden sector model described in Sec.\ref{sec:effpot}
is that the FI term parameter $\lambda$ is 
much smaller than the $SU(2)$ dynamical scale $\Lambda$ and the Landau pole 
$\Lambda_L$
is much larger than the $SU(2)$ dynamical scale,
namely, 
\beq
\lambda^2\ll\Lambda\ll\Lambda_L.
\eeq
This implies that the FI term is treated as perturbation, and that the $U(1)$ gauge coupling is 
always weak in scales below the Landau pole $a_1 < \Lambda_L$ so the $SU(2)$ coupling is 
not affected by the $U(1)$ dynamics. 
Then the analytic properties of the ${\C N}=2$ $SU(2)$ gauge theory is not spoiled by the FI term nor by the $U(1)$ part.

Below we consider only the supersymmetric part of the action 
${\C L}_{\rm SUSY}={\C L}_{\rm VM}+{\C L}_{\rm HM}$, assuming that the FI term is negligible.
We shall focus on the Coulomb branch.
The vector multiplet scalars $a_2$, $a_1$ and their dual variables $a_{2D}$, $a_{1D}$ 
undergo $Sp(4, {\B R})$ duality transformation \cite{Seiberg:1994aj}.
The subgroup of $Sp(4, {\B R})$ that leaves the action invariant is in the form 
\cite{AlvarezGaume:1997fg,AlvarezGaume:1997ek}:
\beq
\left(
\begin{array}{c}
a_{2D} \\ a_{2} \\ a_{1D} \\ a_1
\end{array}
\right)
\mapsto
\left(
\begin{array}{cccc}
\alpha & \beta & 0 & p\\
\gamma & \delta & 0 & q\\
p\gamma-q\alpha & p\delta-q\beta & 1 & -pq\\
0 & 0 & 0 & 1
\end{array}
\right)
\left(
\begin{array}{c}
a_{2D} \\ a_{2} \\ a_{1D} \\ a_1
\end{array}
\right),
\eeq
where
\beq
\left(\begin{array}{cc}\alpha & \beta \\\gamma & \delta\end{array}\right)
\in SL(2,{\B Z})
\eeq
and $p, q\in{\B Q}$. 
A crucial observation here is that the $(a_{2D}, a_2, a_1)$ part of the monodromy transformations is identical to that of the ${\C N}=2$ supersymmetric QCD with gauge group $SU(2)$
and two massive hypermultiplets, where the masses are taken to be $m_1=m_2=\sqrt 2 a_1$. 
This is consistent with our assumption that the $SU(2)$ gauge dynamics is intact. 
The prepotential implied by the monodromy transformation naturally contains the prepotential of
the ${\C N}=2$ supersymmetric QCD \cite{Seiberg:1994aj}
with $N_f=2$ massive hypermultiplets and gauge group $SU(2)$, 
\bea
&&\hspace{-15mm}{\C F}(A_2, A_1, \Lambda, \Lambda_L)\nn\\
&&\hspace{-12mm}={\C F}_{SU(2)}^{\rm SW}(A_2, m_1, m_2, \Lambda)\Big\vert_{m_1=m_2=\sqrt 2 A_1}
+ \half CA_1^2.
\label{eqn:prepot}
\eea
The first term is the supersymmetric QCD part, and the second is a $U(1)$ contribution where
the parameter $C$ is determined by the relative scale between $\Lambda_L$ and $\Lambda$.

The periods $a_2$ and $a_{2D}$ are obtained from the elliptic curve of the 
${\C N}=2$ supersymmetric QCD with two massive hypermultiplets, 
\beq
y^2=\left(x^2-\frac{\Lambda^4}{64}\right)(x-u)+\quarter m_1m_2\Lambda^2 x-\frac{\Lambda^4}{64}
(m_1^2+m_2^2),
\label{eqn:SWcurve}
\eeq
where $u=\Tr A_2^2$, and $m_1=m_2=\sqrt 2 a_1$ now.
The Seiberg-Witten differential is
\beq
\lambda_{\rm SW}=-\frac{\sqrt 2}{4\pi}\frac{y dx}{x^2-\frac{1}{64}\Lambda^4},
\eeq
and the periods are given by the contour integrals,
\beq
a_{2D}=\oint_{\gamma_1}\lambda_{\rm SW}, 
\quad
a_{2}=\oint_{\gamma_2}\lambda_{\rm SW}.
\label{eqn:periods}
\eeq
The cycles $\gamma_1$, $\gamma_2$ are to be specified below.
It is convenient to change the variables as $x=4X+\third u$ and $y=4Y$ so that the
curve (\ref{eqn:SWcurve}) is uniformized in the Weierstrass form,
\beq
Y^2=4X^3-g_2 X-g_3=4(X-e_1)(X-e_2)(X-e_3),
\label{eqn:SWC}
\eeq
with
\bea
g_2&=&\frac{1}{16}\left(\frac 43 u^2+\frac{\Lambda^4}{16}-2a_1^2\Lambda^2\right),\nn\\
g_3&=&\frac{1}{16}\left(\frac{a_1^2\Lambda^4}{16}-\frac{a_1^2\Lambda^2}{6}u
-\frac{\Lambda^4}{96}u+\frac{2}{27}u^3\right),
\eea
and the three roots are
\bea
e_1&=&\frac{u}{24}-\frac{\Lambda^2}{64}
-\eighth\sqrt{(u+\frac{\Lambda^2}{8})^2-2a_1^2\Lambda^2},
\nn\\
e_2&=&\frac{u}{24}-\frac{\Lambda^2}{64}
+\eighth\sqrt{(u+\frac{\Lambda^2}{8})^2-2a_1^2\Lambda^2},
\nn\\
e_3&=&-\frac{u}{12}+\frac{\Lambda^2}{32}.
\label{eqn:roots}
\eea
The three singular points (\ref{eqn:s-f}) arise when two of these roots coincide.
With the change of variables the Seiberg-Witten differential becomes
\beq
\lambda_{\rm SW}=
\frac{\sqrt 2}{4\pi}\frac{dX}{Y}\left(
\frac 23 u-4X-\eighth\frac{a_1^2\Lambda^2}{X-c}\right),
\eeq
with
\beq
c=-\frac{u}{12}-\frac{\Lambda^2}{32}.
\eeq
The contours of the integrals (\ref{eqn:periods}) are fixed by the 
asymptotic behavior of the periods,
\beq
a_{2D}\sim\frac{i}{2\pi}\sqrt{2u}\ln\frac{u}{\Lambda^2},
\quad
a_2\sim\frac{\sqrt{2u}}{2},
\eeq
in the region $u\rightarrow \infty$;
the correct contours turn out to be that 
$\gamma_1$ encircles $\{e_2, e_3\}$ and $\gamma_2$ encircles $\{e_1, e_3\}$.
The periods are now expressed as
\bea
&&\hspace{-10mm}a_{2D}
=\frac{\sqrt 2}{4\pi}\left(
\frac 43uI_1^{(1)}-8I_2^{(1)}-\frac{a_1^2\Lambda^2}{4}I_3^{(1)}(c)
\right),
\label{eqn:a2D}\\
&&\hspace{-10mm} a_{2}
=\frac{\sqrt 2}{4\pi}\left(
\frac 43uI_1^{(2)}-8I_2^{(2)}-\frac{a_1^2\Lambda^2}{4}I_3^{(2)}(c)
\right)+a_1,
\label{eqn:a2}
\eea
using integrals
\bea
&&I_1^{(i)}=\half\oint_{\gamma_i}\frac{dX}{Y},\qquad
I_2^{(i)}=\half\oint_{\gamma_i}\frac{XdX}{Y},\nn\\
&&\qquad I_3^{(i)}(c)=\half\oint_{\gamma_i}\frac{dX}{Y(X-c)}.
\label{eqn:integs}
\eea
It is convenient to express the integrals using the Weierstrass functions through the 
Abel-Jacobi map:
\beq
(\wp(z),\wp'(z))\mapsto(X,Y).
\label{eqn:abel}
\eeq
This is a map from ${\B C}/\Gamma$ to the curve (\ref{eqn:SWC}), where 
$\Gamma$ is the lattice spanned by periods $\omega_1$ and $\omega_2$.
The Weierstrass $\sigma$, $\zeta$, and $\wp$ functions are
\bea
&&\sigma(z)=z\prod_{w\in\Gamma^*}\left(1-\frac zw\right)
\exp\left\{\frac zw+\half\left(\frac zw\right)^2\right\},\nn\\
&&\zeta(z)=\frac{d\ln\sigma(z)}{dz},
\qquad \wp(z)=-\frac{d\zeta(z)}{dz},
\eea
where $\Gamma^*=\Gamma-\{0\}$.
The function $\wp(z)$ is even whereas $\zeta(z)$ and $\sigma(z)$ are odd.
Introducing $\omega_3=-\omega_1-\omega_2$, the half periods are related to the three roots 
(\ref{eqn:roots}) as
$e_i=\wp(\omega_i/2)$, $i=1,2,3$.
Denoting also $\eta_i=\zeta(\frac{\omega_i}{2})$, the following relations are well known:
\bea
&&e_1+e_2+e_3=0,\\
&&\eta_1+\eta_2+\eta_3=0,\\
&&\hspace{-15mm}
\omega_1\eta_2-\omega_2\eta_1=
\omega_2\eta_3-\omega_3\eta_2=
\omega_3\eta_1-\omega_1\eta_3=\pi i.
\label{eqn:Legendre}
\eea
The last identities are known as Legendre's relations.
The following relations are also well known:
\bea
&&\hspace{-10mm}\frac{\wp'(z_0)}{\wp(z)-\wp(z_0)}=2\zeta(z_0)-\zeta(z+z_0)-\zeta(z-z_0),
\label{eqn:pzeta}\\
&&\sigma(z+\omega_i)=-\sigma(z)\exp[\eta_i(2z+\omega_i)].
\label{eqn:sigmaper}
\eea
The inverse of (\ref{eqn:abel}) is
\beq
z=\int_\infty^p\frac{dX}{Y}
=-\frac{1}{\sqrt{e_2-e_1}}F(\phi,k),
\eeq
where
$F(\phi,k)$ is the incomplete elliptic integral of the first kind and 
$k^2=(e_3-e_1)/(e_2-e_1)$, $\sin^2\phi=(e_2-e_1)/(p-e_1)$.
The integrals (\ref{eqn:integs}) are expressed using the Weierstrass functions as
($i=1,2$)
\bea
&&I_{1}^{(i)}=\int_{\frac{\omega_{3-i}}{2}}^{-\frac{\omega_3}{2}}dz=\half\omega_i,\label{eqn:int1}\\
&&I_{2}^{(i)}=\int_{\frac{\omega_{3-i}}{2}}^{-\frac{\omega_3}{2}}\wp(z)dz=-\eta_i,\label{eqn:int2}\\
&&\hspace{-10mm}I_{3}^{(i)}(c)=\int_{\frac{\omega_{3-i}}{2}}^{-\frac{\omega_3}{2}}
\frac{dz}{\wp(z)-\wp(z_0)}
=\frac{\omega_i\zeta(z_0)-2\eta_iz_0}{\wp'(z_0)},\label{eqn:int3}
\eea
where $z_0$ is the point on ${\B C}/\Gamma$ such that
$X=\wp(\pm z_0)=c$, 
$Y=\wp'(\pm z_0)=\pm i a_1\Lambda^2/8\sqrt 2$.


Below we describe a way of computing the matrix
\cite{AlvarezGaume:1997fg,Arai:2001pi,Arai:2007md},
\beq
\tau_{ij}=\frac{\partial^2{\C F}}{\partial a_i\partial a_j}
=\frac{\partial a_{iD}}{\partial a_j}.
\eeq
The computation of $\tau_{22}$ is straightforward.
We have 
\beq
\tau_{22}
=\left.\frac{\partial a_{2D}}{\partial a_2}\right|_{a_1}
=\left.\frac{\partial a_{2D}}{\partial u}\right|_{a_1}\left.\frac{\partial u}{\partial{a_2}}\right|_{a_1}
=\frac{I_1^{(1)}}{I_1^{(2)}}
=\frac{\omega_1}{\omega_2}.
\eeq
Similarly, 
\bea
\tau_{21}
=\left.\frac{\partial a_{2D}}{\partial a_1}\right|_{a_2}
&=&\left.\frac{\partial a_{2D}}{\partial a_1}\right|_{u}
-\tau_{22}\left.\frac{\partial a_{2}}{\partial a_1}\right|_{u}\nn\\
&=&-\frac{\sqrt 2\Lambda^2 a_1}{16\pi}\left(
I_3^{(1)}(c)-\frac{\omega_1}{\omega_2}I_3^{(2)}(c)\right)-\tau_{22}\nn\\
&=&-\frac{2z_0}{\omega_2}-\tau_{22},
\eea
where (\ref{eqn:Legendre}) and (\ref{eqn:int3}) have been used.
For computing $\tau_{11}$ we start with the Euler relation,
\beq
2{\C F}=a_{1}a_{1D}+a_{2}a_{2D}+\Lambda\frac{\partial{\C F}}{\partial\Lambda},
\label{eqn:Euler}
\eeq
which follows from the fact that the ${\C N}=2$ prepotential is a homogeneous function of 
degree 2.
The last term is proportional to the 1-loop beta function
\cite{Matone:1995rx,Sonnenschein:1995hv,Eguchi:1995jh,D'Hoker:1996ph} and in our case
of color $N_c=2$ and flavor $N_f=2$,
\beq
\Lambda\frac{\partial{\C F}}{\partial\Lambda}=\frac{2N_c-N_f}{4\pi i}u=\frac{u}{2\pi i}.
\eeq
Differentiating (\ref{eqn:Euler}) with respect to $a_1$ one obtains,
\beq
a_{1D}=a_1\tau_{11}+a_2\tau_{21}-\frac{1}{2\pi i}\left.\frac{\partial u}{\partial a_2}\right\vert_{a_1}\left.\frac{\partial a_2}{\partial a_1}\right\vert_u.
\label{eqn:a1DEuler}
\eeq
On the right-hand side, 
\bea
\left.\frac{\partial u}{\partial a_2}\right\vert_{a_1}
&=&\left(
\left.\frac{\partial a_2}{\partial u}\right\vert_{a_1}\right)^{-1}
=\frac{8\pi}{\sqrt 2\omega_2},\\
\left.\frac{\partial a_2}{\partial a_1}\right\vert_{u}
&=&-\frac{\sqrt 2 a_1\Lambda^2}{16\pi}I_3^{(2)}(c)+1,
\eea
and $a_2$ is (\ref{eqn:a2}) with the integrals given by 
(\ref{eqn:int1}), (\ref{eqn:int2}), (\ref{eqn:int3}).
In order to evaluate $a_{1D}$ on the left-hand side, we use the reciprocity law
\cite{GriffithsHarris:1994}
for differentials of the first kind (a holomorphic 1-form) $\chi$, and the third kind (a meromorphic 
form with single poles) $\psi$, on a genus one Riemann surface,
\beq
\oint_{\gamma_2}\chi\oint_{\gamma_1}\psi-\oint_{\gamma_1}\chi\oint_{\gamma_2}\psi
=2\pi i\sum_{n}\left({\rm Res}_{x_n^+}\psi\right)\int_{x_n^-}^{x_n^+}\chi,
\eeq
where $x_n^+$ and $x_n^-$ are the poles on the positive and negative Riemann sheets.
Applying this to $\chi=\partial\lambda_{\rm SW}/\partial a_2\vert_{a_1}$
and 
$\psi=\partial\lambda_{\rm SW}/\partial a_1\vert_u$, we have
\bea
\left.\frac{\partial a_{1D}}{\partial a_2}\right\vert_{a_1}
&=&\left.\frac{\partial a_{2D}}{\partial a_1}\right|_{u}
-\left.\frac{\partial a_{2D}}{\partial a_2}\right\vert_{a_1}
\left.\frac{\partial a_{2}}{\partial a_1}\right|_{u}\nn\\
&=&-\int_{x_0^-}^{x_0^+}\left.\frac{\partial\lambda_{SW}}{\partial a_2}\right\vert_{a_1},
\eea
where $x_0^\pm=\wp(\pm z_0)$ on the two Riemann sheets for the point $X=c$.
Integrating, we obtain
\beq
a_{1D}=-\int_{x_0^-}^{x_0^+}\lambda_{SW}+\tilde C(a_1),
\eeq
with $\tilde C(a_1)$ the constant of integration. 
The integral can be expressed using the Weierstrass functions as
\bea
\int_{x_0^-}^{x_0^+}\frac{dX}{Y}&=&\int_{-z_0}^{z_0}dz=2z_0,\\
\int_{x_0^-}^{x_0^+}\frac{XdX}{Y}&=&\int_{-z_0}^{z_0}\wp(z)dz=-2\zeta(z_0),\\
\int_{x_0^-}^{x_0^+}\frac{dX}{Y(X-c)}&=&\int_{-z_0-\epsilon}^{z_0+\epsilon}
\frac{dz}{\wp(z)-\wp(z_0)}\nn\\
&&\hspace{-25mm}=
\frac{2}{\wp'(z_0)}\left[
2z_0\zeta(z_0)-\ln\sigma(2z_0)+\ln\sigma(\epsilon)\right],
\eea
where the last integral is divergent and the regulator $\epsilon$ has been introduced. 
It is possible to pass on the divergence to the integration constant and $\tilde C(a)$ is
now related to $C$ of (\ref{eqn:prepot}) as
$\tilde C(a_1)=a_1(C+\frac{i}{\pi}\ln\sigma(\epsilon))$.
Inserting these expressions into (\ref{eqn:a1DEuler}) we finally obtain,
\bea
\tau_{11}=
\frac{i}{\pi}\left(
\ln\sigma(2z_0)-4z_0^2\frac{\eta_2}{\omega_2}\right)-2\tau_{21}-\tau_{22}+C.
\eea
In the numerical computations we have set $C=8\pi i$ which corresponds to $\Lambda_L/\Lambda\sim 10^{17-18}$.

\section{The ${\C N}=2$ Fayet-Iliopoulos term}

In ${\cal N}=1$ supersymmetric theory, the FI term is written as the D-component of a vector 
superfield. 
With extended ${\cal N}=2$ SUSY, in contrast, the FI term can also be written as
the F-component of a chiral superfield, forming a triplet under the $SU(2)_R$ symmetry.
In the main text of the paper we used the F-component form of the FI term.
In this Appendix we show that these two forms have the same origin in ${\cal N}=2$ theory,
and in fact are equivalent up to the $SU(2)_R$ symmetry.

We start with a review of the ${\cal N}=2$ projective superspace formalism 
\cite{Karlhede:1984vr,Lindstrom:1987ks,Lindstrom:1989ne,PSS},
which is convenient for our purpose as we wish to decompose 
the fields into ${\cal N}=1$. 
For an alternative harmonic superspace formalism, the reader is referred to a nice review
\cite{Galperin:2001uw} and references therein.
It is known that the standard superspace extended to ${\cal N}=2$,
$z^M\equiv(x^\mu,\theta^{i\alpha},\bar{\theta}^i_{\dot{\alpha}})$, 
$\mu=0,1,2,3$ and $i=1,2$, fails to describe the off-shell interactions.
In projective superspace an extra complex bosonic coordinate $\zeta$ is introduced, 
supplementing the standard $z\equiv z^M$.
Here, $\zeta$ is called the projective coordinate and parameterizes 
${\B C}{\B P}^1=SU(2)_R/U(1)$.

The ${\C N}=2$ supercovariant derivatives satisfy the following algebra,
\begin{eqnarray}
&&\{D_{i\alpha}, D_{j\beta}\}
 =\{\overline{D}_{i\dot{\alpha}}, \overline{D}_{j\dot{\beta}}\}=0, \nn\\
&&\{D_{i\alpha}, \overline{D}^j_{\dot{\beta}}\}
 =-2i\delta^j_i\sigma^\mu_{\alpha\dot{\beta}}\partial_\mu \,.
\end{eqnarray}
In an Abelian subspace of ${\C N}=2$ superspace parametrized by $\zeta$, 
the supercovariant derivatives are defined as,
\bea
\nabla_{\alpha}(\zeta)&=&D_{1\alpha}+\zeta D_{2\alpha} \, ,\label{eqn:sucovderiv1}\\
\overline{\nabla}_{\dot{\alpha}}(\zeta)&=&\overline{D}^2_{\dot{\alpha}}
 -\zeta \overline{D}^1_{\dot{\alpha}} \, .\label{eqn:sucovderiv2}
\eea
Note that the conjugate of an operator is given by Hermitian conjugation combined with 
the antipodal map on the Riemann sphere $\bar\zeta\rightarrow-1/\zeta$, 
up to a multiplicative factor.
For example,
\beq
\overline{\nabla}_{\dot\alpha}(\zeta)
=(-\zeta)\overline{\nabla_\alpha(-\frac{1}{\zeta})}
=(-\zeta)\left(\overline{D}^1_{\dot\alpha}-\frac{1}{\zeta}\overline{D}^2_{\dot\alpha}\right),
\eeq
yielding (\ref{eqn:sucovderiv2}).
Superfields in the projective superspace are bundles on the subspace of $(z,\zeta)$
satisfying constraints (analogous to chiral superfields in ${\cal N}=1$ superspace),
\beq
\nabla_\alpha \Upsilon(z,\zeta)
 =\overline{\nabla}_{\dot{\alpha}}\Upsilon(z,\zeta)=0\, .\label{const1}
\eeq
The restricted measure of the subspace is constructed from operators
$\Delta_\alpha\equiv \zeta^{-1}D_{1\alpha}-D_{2\alpha}$, 
$\overline{\Delta}_{\dot\alpha}\equiv
\zeta^{-1}\overline{D}^2_{\dot\alpha}+\overline{D}^1_{\dot\alpha}$,
which are orthogonal to (\ref{eqn:sucovderiv1}), (\ref{eqn:sucovderiv2}).
A manifestly ${\cal N}=2$ invariant action in projective superspace is given as an integral
of a real functional of superfields using this measure,
$
\oint_C\zeta d\zeta\Delta^2\overline{\Delta}{}^2,
$
with some contour $C$ in the $\zeta$-plane.
Setting $\theta\equiv\theta_1$, the measure reduces to the standard ${\cal N}=1$ 
Grassmann measure $\int d^4\theta$, and the action is written in the ${\cal N}=1$ superspace.

In projective superspace the ${\cal N}=2$ invariant FI term takes the form,
\begin{eqnarray}
{\cal L}_{\rm FI}=\oint_C {d\zeta \over 2\pi i \zeta}\int d^4\theta \xi(\zeta) {\cal V}(z,\zeta)
, \label{FI-p}
\end{eqnarray}
where $\xi(\zeta)$ is the coefficient of the FI term given by
\begin{eqnarray}
\xi(\zeta)=\frac{\lambda}{\zeta}+\rho-\zeta \bar\lambda,
\quad\rho \in {\B R},
\quad\lambda \in {\B C},
\label{coef}
\end{eqnarray}
and ${\cal V}$ is a so-called real tropical multiplet, satisfying the constraints (\ref{const1}) 
and is expanded in $\zeta$ as
\begin{eqnarray}
{\cal V}(z,\zeta)=\sum_{n=-\infty}^\infty v_n(z) \zeta^n, \label{tor}
\end{eqnarray}
where $v_n$'s are ${\cal N}=1$ superfields. 
From (\ref{const1}) it follows that, 
\begin{eqnarray}
D_{1\alpha} v_{n+1} = -D_{2\alpha} v_n,~~~\bar{D}^1_{\dot{\alpha}}v_n = \bar{D}^2_{\dot{\alpha}}v_{n+1},
\end{eqnarray}
and the reality condition ${\cal V}=\bar{\cal V}$ constrains
$
 v_{-n}=(-)^n\bar{v}_n. \label{conj}
$
Among $v_n$, all but $v_{-1}, v_0$ and $v_{1}$ are gauged away by the $U(1)$ gauge transformation \cite{Lindstrom:1989ne}.
The superfield $v_1$ is a prepotential of a chiral superfield, and $v_0$ is a vector superfield.

Substituting (\ref{coef}) and (\ref{tor}) into (\ref{FI-p}), we have
\begin{eqnarray}
{\cal L}_{\rm FI}&=&\int d^4\theta (\bar{\lambda}\bar{v}_{1}+\rho v_0 + \lambda v_1)
\nn\\
&=&\rho \int d^4\theta V + \left(\lambda \int d^2\theta A + h.c.\right), \label{LFI}
\end{eqnarray}
where $V\equiv v_0$ is a vector superfield and $A\equiv -\bar{D}^2 v_1/4$ is a chiral superfield.
We have used $d^2\theta \sim -D^2/4$, $d^2\bar{\theta}\sim-\bar{D}^2/4$ to go from the first to the
second line.

We may rewrite (\ref{LFI}) in a way that the $SU(2)_R$ symmetry becomes manifest:
\beq
{\cal L}_{\rm FI} 
=\Tr (\Xi {\C D}),
\eeq
%
where $\Xi=\Xi^{ij}$ and ${\cal D}={\cal D}_{ij}$ with $i, j$ the $SU(2)_R$
indices,
\begin{eqnarray}
\Xi^{ij}&=&i(\sum_{a=1}^3\sigma^a \xi^a \epsilon)^{ij},\quad
\epsilon=\left(\begin{array}{cc}
0 & 1 \\
-1 & 0
\end{array}
\right),\\
{\C D}_{ij}&=& \left(
\begin{array}{cc}
 F_A & -iD_V \\
 -iD_V & \bar{F_A}
\end{array}
\right).
\end{eqnarray}
Here $\sigma^a$ are the Pauli matrices and $F_A, D_V$ are the auxiliary fields of $A$ and $V$.
The coefficients of the $SU(2)_R$ rotation are
\begin{eqnarray}
\xi^1=-{\lambda-\bar{\lambda} \over 2i},~~~\xi^2=-{\lambda+\bar{\lambda} \over 2},~~~
\xi^3=\frac{\rho}{4}.
\end{eqnarray}
It is now obvious that by using the $SU(2)_R$ symmetry, we can take $\rho=0$ or 
$\lambda=0$. In our model, we have chosen $\rho=0$.

\end{appendix}


\end{document}